\documentclass[a4paper,11pt]{article}

\usepackage{adjustbox}
\usepackage{amsfonts}
\usepackage{amssymb, amscd}
\usepackage{graphicx}
\usepackage{empheq}
\usepackage{mathrsfs}
\usepackage[table, svgnames, dvipsnames, xcdraw]{xcolor}
\usepackage{colortbl}
\usepackage{slashed}
\usepackage{pdfpages}
\usepackage{braket}
\usepackage{sidecap}
\usepackage{multirow}
\usepackage{adjustbox}
\usepackage{arydshln}
\usepackage[font=small,labelsep=none]{caption}
\usepackage{mathtools}
\usepackage{cite}
\usepackage{hyperref}
\usepackage{tikz}
\usetikzlibrary{arrows,decorations.markings}
\usepackage{subcaption}
\usepackage[mode=buildnew]{standalone}
\usepackage[normalem]{ulem}
\usepackage[left=2cm,right=2cm,top=3.5cm,bottom=3.5cm]{geometry}

\hypersetup{
    colorlinks=true,
    linkcolor=ceruleanblue,
    filecolor=ceruleanblue,      
    urlcolor=ceruleanblue,
    citecolor=ceruleanblue,
}

\definecolor{myblue}{RGB}{174, 198, 219}
\definecolor{myred}{RGB}{157,31,68}
\definecolor{ceruleanblue}{rgb}{0.0, 0.2, 0.6}
\definecolor{grey}{rgb}{0.4,0.4,0.4}
\definecolor{dullmagenta}{rgb}{0.4,0,0.4}
\definecolor{darkblue}{rgb}{0,0,0.4}
\definecolor{midblue}{rgb}{0,0,0.5}
\definecolor{midred}{rgb}{0.5,0,0}
\definecolor{orange}{rgb}{1,0.5,0}
\definecolor{lightbrown}{rgb}{0.75,0.5,0.25}
\definecolor{tan}{cmyk}{0.14,0.42,0.56,0}
\definecolor{djunglegreen}{cmyk}{0.99,0,0.52,0}
\definecolor{lightgreen}{rgb}{0,1,0}
\definecolor{olivegreen}{cmyk}{0.64,0,0.95,0.40}
\definecolor{midgreen}{rgb}{0.0,0.675,0.0}
\definecolor{darkgreen}{rgb}{0,0.5,0}
\definecolor{ceruleanblue}{rgb}{0.0, 0.2, 0.7}
\definecolor{burgundy}{rgb}{0.5, 0.0, 0.13}
\definecolor{blue_light}{RGB}{0, 102, 204}

\newlength\mytemplen
\newsavebox\mytempbox

\makeatletter
\newcommand\mybox{%
    \@ifnextchar[
       {\@mybox}%
       {\@mybox[0pt]}}

\def\@mybox[#1]{%
    \@ifnextchar[
       {\@@mybox[#1]}%
       {\@@mybox[#1][0pt]}}

\def\@@mybox[#1][#2]#3{
    \sbox\mytempbox{#3}%
    \mytemplen\ht\mytempbox
    \advance\mytemplen #1\relax
    \ht\mytempbox\mytemplen
    \mytemplen\dp\mytempbox
    \advance\mytemplen #2\relax
    \dp\mytempbox\mytemplen
    \colorbox{black!10!white}{\hspace{1em}\usebox{\mytempbox}\hspace{1em}}}

\def\thickhline{\noalign{\hrule height.8pt}}

\def\MP{M_{\rm Pl}}
\newcommand{\vect}[1]{\boldsymbol{#1}}
\newcommand{\de}{\, \mathrm{d}}	 

\numberwithin{equation}{section}
\newcommand{\abs}[1]{\lvert#1\rvert}

\usepackage{multicol}

\usepackage{titlesec}

\tikzset{cross/.style={cross out, draw=black, minimum size=2*(#1-\pgflinewidth), inner sep=0pt, outer sep=0pt},
cross/.default={4pt}}

\tikzset{
    master/.style={
        execute at end picture={
            \coordinate (lower right) at (current bounding box.south east);
            \coordinate (upper left) at (current bounding box.north west);
        }
    },
    slave/.style={
        execute at end picture={
            \pgfresetboundingbox
            \path (upper left) rectangle (lower right);
        }
    }
}

\date{\today}

\begin{document}
	
	\begin{center}
		\LARGE{\bf To the Problem of Cosmic Expansion in Massive Gravity}
		\\
        \vspace{.5cm}
		
		\large{Lavinia Heisenberg$^{\,\rm a}$, Alessandro Longo$^{\,\rm b, c}$, Giovanni Tambalo$^{\,\rm d}$ \\ and Miguel Zumalacarregui$^{\,\rm e}$}
		\\
        \vspace{.5cm}
		
		\small{
			\textit{$^{\rm a}$
				Institut f\"ur Theoretische Physik, Philosophenweg 19, 69120 Heidelberg, Germany}}
		\vspace{.2cm}
		
		\small{
			\textit{$^{\rm b}$
				SISSA, International School for Advanced Studies, 34136 Trieste, Italy}}
		\vspace{.2cm}
		
		\small{
			\textit{$^{\rm c}$
				INFN, National Institute for Nuclear Physics, 34127 Trieste, Italy}}
		\vspace{.2cm}
		
		\small{
			\textit{$^{\rm d}$
				Institut f\"ur Theoretische Physik, ETH Z\"urich, 8093 Z\"urich, Switzerland}}
		\vspace{.2cm}
		
		\small{
			\textit{$^{\rm e}$
				Max Planck Institute for Gravitational Physics (Albert Einstein Institute) \\
Am Mühlenberg 1, D-14476 Potsdam-Golm, Germany}}
		\vspace{.2cm}
				
	\end{center}
	
	\vspace{0.3cm} 
	
	\begin{abstract}
    \normalsize
We consider evolving, spatially flat isotropic and homogeneous (FLRW) cosmologies in ghost-free (dRGT) massive gravity. In this theory, no dynamical flat FLRW background exists if the reference metric is chosen to be Minkowski and the Stueckelberg fields are homogeneous. Relaxing the assumptions on the Stueckelberg profiles gives access to dynamical backgrounds. We propose a classification of the viable flat FLRW cosmological solutions of dRGT massive gravity.
Instead of specifying an initial ansatz for the Stueckelberg fields $\phi^a$ and the reference metric $f_{ab}$, we show that imposing homogeneity and isotropy on the square root tensor $X^{\mu}_{\nu}=\left(\sqrt{g^{-1}\partial\phi^a \partial\phi^bf_{ab}}\right)^{\mu}_{\nu}$
leads to dynamical cosmological solutions, and we characterize their properties. These solutions become dynamical only when the Stueckelberg fields acquire a sufficiently inhomogeneous and/or anisotropic profile. 
We explore the consequences for the minimal model and the complete dRGT theory, and show that perturbations are strongly coupled, at the quadratic level, on these backgrounds.
	\end{abstract}
	\vspace{2cm}
	
	\newpage
	{
		\hypersetup{linkcolor=black}
		\setcounter{tocdepth}{2}
		\tableofcontents
	}

	\flushbottom
	
	\vspace{1cm}

\section{Introduction}

Einstein's General Relativity (GR) is the unique theory that describes the nonlinear physics of a massless spin-2 field, identified with the metric tensor $g_{\mu \nu}$ of a Lorentzian manifold. Its phenomenology is so vast that many of its predictions have just been recently confirmed (e.g.~mapping the spacetime around supermassive Black Holes~\cite{EventHorizonTelescope:2019dse} or recording Gravitational Waves \cite{PhysRevLett.116.061102}).
Furthermore, GR is an essential pillar of the standard cosmological model, describing the evolution of the universe over 13.8 billion years~\cite{Planck:2018vyg}.

Despite its great success in describing data, GR requires the introduction of cold dark matter (CDM) to explain structure formation~\cite{Bertone:2004pz} and a cosmological constant ($\Lambda$) to explain the late-time cosmic acceleration~\cite{Weinberg:2013agg}. In the $\Lambda$CDM model, the dark components make up $95\%$ of the universe's energy.
The observed late-time cosmic acceleration motivates the study of \textit{dark energy} models different than the constant $\Lambda$, including gravity theories beyond GR.
These investigations are further motivated by growing tensions between disparate datasets interpreted in the simple $\Lambda$CDM model~\cite{Abdalla:2022yfr,DESI:2024mwx}.

A wide range of theories have been proposed and explored~\cite{Clifton:2011jh,Ishak:2018his,CANTATA:2021asi}. Among them,
Massive Gravity has gained a renewed interest in the last decade, when the construction
of the de Rham-Gabadadze-Tolley (dRGT) model \cite{de_Rham_2010,deRham:2010kj,deRham:2014zqa,Hinterbichler_2012} showed how to overcome
the technical difficulties that the introduction of a mass term for the graviton brings. Many applications of the theory were explored through the years, from the study of Black Holes solutions \cite{Volkov_2013} to the extraction of ``positivity bounds" on the dRGT parameters in the flat spacetime setting\cite{Bellazzini_2018,Bellazzini_2024}. However, exploring its cosmological implications has been made extremely challenging by the absence of a dynamical flat FLRW background \cite{DAmico:2011eto}. More precisely, the choice of a Minkowski reference metric, together with an FLRW-compatible ansatz for the Stueckelberg fields $\phi^a$, results in a staticity constraint on the scale factor
\begin{equation}
\dot{a}=0 \ .
\end{equation}
Progress was made (see \cite{DAmico:2012qbv,Volkov:2012zb}) when it was realized that the aforementioned No-Go theorem was a consequence of the poorly chosen ansatz for the Stueckelberg fields. Indeed, different authors showed that relaxing homogeneity or isotropy \cite{Volkov:2012zb, G_mr_k_o_lu_2012, Gratia_2012} of the Stueckelbergs gave access to dynamical FLRW backgrounds, even when the extra metric $f_{ab}$ was chosen to be Minkowski. Among them, a potentially interesting self-accelerating solution, in which the Stueckelberg sector behaves as a cosmological constant.
It was later highlighted that, on this background, three of the five degrees of freedom of the theory are strongly coupled (see \cite{DAmico:2012qbv,Tasinato_2013e,Khosravi_2013,Tasinato_2013v,Felice_2013}). 
Despite its pathologies, the extraction of the self-accelerating background teaches an interesting lesson: inhomogeneous Stueckelberg configurations can behave as an effective perfect fluid and sustain a dynamical flat FLRW background solution. 

This paper exploits and extends the approach proposed in \cite{Volkov:2012zb,Y_lmaz_2014_1} to classify the most general dynamical flat FLRW solutions of dRGT massive gravity.
Starting from the assumption that the building block of the theory, $X^{\mu}_{\nu}$, enjoys the FLRW symmetries of the physical metric, we find that the admissible dynamical backgrounds are the aforementioned self-accelerated solution and a scenario in which the Stueckelberg stress-energy tensor behaves as a mixture of perfect fluids. A similar idea was presented by \cite{Y_lmaz_2014_1}\footnote{Unfortunately, their proposed solutions for the profiles of the Stueckelberg fields leads again to the aforementioned No-Go theorem.}, and by \cite{Volkov:2012zb}, under the assumption of a diagonal reference metric and of spherically symmetric configurations of the Stueckelberg fields.

We advance these results and show that the existence of these solutions is actually guaranteed for any possible choice of $f_{ab}$ and $\phi^a$ which satisfy a set of consistency conditions (in the form of complicated partial differential equations). These consistency relations arise as a consequence of the symmetries imposed on $X^{\mu}_{\nu}$ and they can be used, once a choice of $f_{ab}$ is specified, to determine the available profiles of $\phi^a$. We explore the perturbative stability of the solutions in the absence and in the presence of additional matter content. 

The paper is organized as follows. Section \ref{SecPX} motivates the methodology in the context of a multi-scalar theory. 
The concrete application of the proposed approach is then implemented and discussed in Sec.~\ref{sec:dRGT_bkg} for the minimal and full theory of dRGT Massive Gravity. Finally, the perturbative stability of the relevant backgrounds, in the absence of matter, is explored in Sec.~\ref{sec:perturbs}. 

Appendix \ref{appconfig} is devoted to the classification of the admissible Stueckelberg profiles, with a focus on the case $f_{ab}=\eta_{ab}$; while App.~\ref{apppert} extends the perturbative analysis in the presence of additional matter.

\section{Single-field \texorpdfstring{$P(X)$}{P(X)} vs multi-field \texorpdfstring{$P(\mathcal{X})$}{P(X)}}\label{SecPX}
As a motivating warm-up, let us study the following toy model: a $P(X)$ theory of a shift-symmetric scalar field minimally coupled to gravity. The action takes the compact form\begin{equation}
	S = \int \de ^4x \sqrt{-g}P(X) \ .
\end{equation}
with $X=-(\partial \phi)^2 / 2$ the canonical kinetic term of the scalar field. Varying the above action with respect to the metric tensor yields the stress-energy tensor of the field $\phi$:
\begin{equation}
	T_{\mu \nu} 
	\equiv 
	\frac{-2}{\sqrt{-g}}\frac{\delta S}{\delta g^{\mu \nu}}
	= 
	P_X \partial_{\mu}\phi \partial_{\nu}\phi +g_{\mu \nu}P  
	\ ,
\end{equation}
where $P_X$ is the derivative of $P$ with respect to $X$.
The above expression tells us that $\phi$ behaves as a fluid with 4-velocity $u^{\mu}\propto \partial^{\mu}\phi$. If we want to do cosmology it is convenient to go to the comoving frame of the fluid and project Einstein equations along the direction of $u^{\mu}$ as well as on the directions orthogonal to the flow. It is clear that $T_{\mu \nu}$ will inherit the symmetries of the spacetime it lives in. This means that if we demand our spacetime to be of an FLRW type (say $g_{00}=-1\>,\> g_{ij}=a(t)^2 h_{ij}$), the matter content shall enjoy the symmetries of an FLRW universe, which are homogeneity and isotropy. More concretely, for a flat FLRW background, these requirements impose the following restrictions on the stress-energy tensor
\begin{equation}
	T_{00}=\rho(t) 
	\ , \quad
	T_{0i}=0
	\ , \quad
	T_{ij}=a(t)^2 p(t)\delta_{ij}
	\ ,
\end{equation}
where $\rho(t)$ and $p(t)$ are the energy density and pressure, respectively, of the scalar field $\phi$.

The question we want to address now is the following: what are the possible field configurations $\phi(t, \vect x)$ for which $T_{\mu \nu}$ enjoys the FLRW symmetries? Together with the homogeneity of $P(X)$, compatibility with the FLRW symmetries would require $T_{0i}=0$, which ultimately strictly forces
\begin{equation}
	\partial_i \phi=0 \ .
\end{equation}
Therefore, it can be concluded that, at least in a single-field model, only homogeneous field configurations behave as a perfect fluid. In particular, homogeneity of $X$ is implied by FLRW compatibility of $T_{\mu \nu}$.\\
We now show that the situation can change as soon as multiple fields are included in the description: in the context of multi-field $P(X)$-like theories, inhomogeneous field configurations can effectively behave as a perfect fluid. To this end, let us take a collection of shift-symmetric scalar fields $\phi^a$ and replace the old kinetic term $X$ with its multi-field version
\begin{equation}
	\mathcal{X}
	=
	-\frac{1}{2}\partial^{\mu}\phi^a \partial_{\mu}\phi^b f_{ab} \ ,
\end{equation}
where $f_{ab}$ is the metric in field space. The Lagrangian density is now given by $P(\mathcal{X})$ and the stress-energy tensor acquires the form
\begin{equation}
	T_{\mu \nu}
	=
	P_{\mathcal{X}}\partial_{\mu}\phi^a \partial_{\nu}\phi^bf_{ab}+g_{\mu \nu}P
	\ .
\end{equation}
It is interesting to look at how the previous constraints of homogeneity and isotropy change with the addition of extra fields:
\begin{displaymath}
\begin{aligned}
    T_{00} \ :\quad  & P_{\mathcal{X}}\dot{\phi}^a\dot{\phi}^bf_{ab}-P =\rho	\ ,\\
    T_{0i}=0 \ : \quad & \partial_i \phi^a \dot \phi^b f_{ab}  =0  \ , \\
    T_{ij}\ :\quad & P_{\mathcal{X}}\partial_i \phi^a \partial_j \phi^b f_{ab}+a^2\delta_{ij}P  =a^2 p\delta_{ij} \ .
\end{aligned}
\end{displaymath}
The very presence of extra fields gives the chance to avoid the strict condition $\partial_i \phi$ that we hit previously. Of course, $\partial_i \phi^a=0\>$ for all $a$ satisfies the above constraints.
However, having multiple fields leaves room, in principle, for inhomogeneous field configurations to satisfy the above system of equations.
When this is the case, we are facing the peculiar situation in which the collection of fields, despite being inhomogeneous, effectively behaves as a perfect fluid (see \cite{Endlich_2013}). 

A concrete example in a $1+1$ setting with a constant diagonal field-space metric is given by the configurations
\begin{equation}
    \phi^a=T^a(t)+A^a+B^a x 
    \>,\quad 
    \dot{T}^0=-\dot{T}^1\frac{B^1 f_{11}}{B^0 f_{00}}
    \ , 
\end{equation}
where $A^a$ and $B^a$ are contants. It is straightforward to check that the above choice of profiles yields a perfect fluid stress-energy tensor. The continuity equation
\begin{equation}
    \partial_t (P_{\mathcal{X}}a \dot{T}^1)=0
     \ ,
\end{equation}
can then be used to determine $T^1(t)$, while the equation of state of the fluid is specified by the choice of $P(\mathcal{X})$. The careful reader will surely have noticed the formal similarity between the scalar $\mathcal{X}$ and the trace of the tensor $(X^2)^{\mu}_{\nu}$ appearing in the dRGT action. Indeed, the Stueckelberg fields $\phi^a$ are not just scalars under diffeomorphism, they also enjoy a global shift symmetry
\begin{equation}
\phi^a\to \phi^a + C^a \ , 
\end{equation}
being the Goldstone bosons of broken diffeomorphisms. This similarity motivates the implementation of the previous approach within the context of dRGT Massive Gravity: instead of starting by specifying an ansatz for the Stueckelberg fields and the reference metric we will impose the FLRW symmetries directly on $X^{\mu}_{\nu}$ (and, as a consequence, on the tensor $(X^2)^{\mu}_{\nu}$), extracting some interesting cosmological predictions which will not rely (up to a certain degree) on a specific choice of $\phi^a$ and $f_{ab}$.
The basic assumption that will be kept through the whole discussion is that there exist at least a choice of $f_{ab}$ and an associated family of configurations $\phi^a$ that make $(X^2)^{\mu}_{\nu}$ compatible with the FLRW symmetries. We will address the task of finding nontrivial configurations, in the particular case of Minkowski reference metric, in App.~\ref{appconfig}. For the time being, let us take homogeneity and isotropy of $X^{\mu}_{\nu}$ as our starting point and explore the cosmological consequences in the framework of dRGT.

\section{Applications to dRGT}\label{sec:dRGT_bkg}

\subsection{dRGT: a quick recap}

As the name suggests, Massive Gravity is an infrared modification of GR in which the graviton is given a mass. Although the introduction of a mass term is generally accompanied by the presence of ghosts, dRGT Massive Gravity \cite{deRham:2010kj,deRham:2014zqa} evades such difficulties by constructing the graviton potential in a very specific way. To be more concrete, together with the physical metric $g_{\mu \nu}$ and the Stueckelberg fields $\phi^a$, an extra (reference) metric $f_{ab}$ is needed in order to build the square root tensor 
\begin{equation}
	\label{eq:squared}X^{\mu}_{\nu}\>,
	\quad
	\mathrm{such \;that}
	\quad
	(X^2)^{\mu}_{\nu}=g^{\mu \rho} \partial_{\rho}\phi^a \partial_{\nu}\phi^b f_{ab} \ .
\end{equation}The action then acquires the form
\begin{equation}\label{eq:dRGT_action}
    S_{\rm dRGT}
    =
    \int \de ^4x \sqrt{-g}
    \left[R-\frac{m^2}{2}
    \sum_{i = 2}^4 \alpha_i \ \mathcal{U}_i[\mathcal{K}]\right] \ ,
\end{equation}
where the dRGT potentials $\mathcal{U}_i[\mathcal{K}]$ are the elementary symmetric polynomials of the tensor
$\mathcal{K}^{\mu}_{\nu} \equiv \delta^{\mu}_{\nu}-X^{\mu}_{\nu}$, namely
\begin{equation}\label{eq:dRGT_potentials}
    \begin{aligned}
        &	\mathcal{U}_2   = [\mathcal{K}^2]-[\mathcal{K}]^2 \ , \\
        &	\mathcal{U}_3	= -2[\mathcal{K}^3]+3 [\mathcal{K}][\mathcal{K}^2]-[\mathcal{K}]^3 \ , \\
		&	\mathcal{U}_4   = 6 [\mathcal{K}^4]-3 [\mathcal{K}^2]^2+ 6 [\mathcal{K}^2][\mathcal{K}]^2-8 [\mathcal{K}][\mathcal{K}^3]-[\mathcal{K}]^4 \ .
    \end{aligned}
\end{equation}
where $m$ is the graviton mass, the $\alpha_2$, $\alpha_3$, $\alpha_4$ are dimensionless coefficients\footnote{Notice that a rescaling of the $\alpha_i$ is degenerate with a rescaling of the mass and therefore one is always free to set $\alpha_2 = 1$, for example.} 
and $[ \mathcal K^n ]$ stands for the trace of the matrix $\mathcal K^n$. For the present discussion, we specialize to a flat FLRW physical metric $g_{\mu\nu}$, which, in comoving coordinates, takes the form $\de s^2 = - \de t^2 + a(t)^2 \de \vect x^2$. As previously stated, we impose homogeneity and isotropy on the tensor $X^{\mu}_{\nu}$, which is the building block of the dRGT action. This means that
\begin{equation}\label{eq:sqroot}
	X^0_0
	=
	F(t) 
	\ ,\quad 
	X^0_i
	=
	0
	\ ,\quad 
	X^i_j
	=
	G(t) \, \delta_{ij}
	\ ,
\end{equation}
where $\delta_{ij}$ 
is the Kronecker delta and $F(t)$ and $G(t)$ are generic functions of time.
As a consequence, both $(X^2)^{\mu}_{\nu}$ and the stress-energy tensor of the Stueckelberg sector enjoys the same symmetries of $X^{\mu}_{\nu}$. Notice that, in general, the contrary is not true: an FLRW-compatible stress tensor for $\phi^a$ does not imply that $X^{\mu}_{\nu}$ is of the form \eqref{eq:sqroot}.
Equation \eqref{eq:squared} results in a set of constraints on the Stueckelberg fields
\begin{equation}\label{eq:constraintsstuck}
		F^2 
		= -\dot{\phi}^a\dot{\phi}^bf_{ab}			\ , \quad 
		0
		=
		\dot{\phi}^a\partial_i \phi^b f_{ab}		\ , \quad 
		G^2\delta_{ij}
		=
		\frac{1}{a^2}\partial_i \phi^a \partial_j \phi^b f_{ab} \ .
\end{equation}
Combining the above system with the Euler-Lagrange equations of the Stueckelberg fields would enable us to determine $\phi^a$ in terms of $F$ and $G$, identifying the family of configurations that support a dynamical cosmology and, at the same time, ensures the symmetry conditions \eqref{eq:sqroot}. 
This step is however not strictly needed. Indeed, we are interested in the background cosmological evolution and this is determined by the behaviour of $F(t)$ and $G(t)$, the components of $X^\mu_\nu$.
To proceed in this direction, we will need the trace of $X^{\mu}_{\nu}$:
\begin{equation}
    [X]=F+3G \;.
\end{equation}
Before moving on, we note that imposing the form \eqref{eq:sqroot} for the building block $X^\mu_\nu$ implies that spatial translations and rotations are unbroken at the level of the perturbations. Indeed, any variation of the action is proportional to traces of $X^\mu_\nu$ (and its powers) thus the inhomogeneous background components of $\phi^a$ do not appear directly. We will see this explicitly in Sec.~\ref{sec:perturbs}. 
On the other hand, imposing FRLW symmetries directly on $T_{\mu\nu}$ would not be sufficient to preserve these symmetries at the level of perturbations.

\subsection{dRGT: Minimal Model}\label{sub:Minimal}
Remembering that $\mathcal{K}^{\mu}_{\nu}=\delta^{\mu}_{\nu}-X^{\mu}_{\nu}$, the explicit form of the minimal potential is
\begin{equation}
    \mathcal{U}_2
    =
    [\mathcal{K}^2]-[\mathcal{K}]^2
    =
    -12 +6[X]-[X]^2+[X^2]
    =
    -6\left(2-F-3G+ G^2+FG\right)
    \ .
\end{equation}
In the presence of the dRGT potential, the Einstein equations are modified as
\begin{equation}
    G_{\mu \nu}-\frac{m^2}{2}Y_{\mu \nu}
    = 
    \frac{T_{\mu \nu}}{\MP^2}
    \ ,
\end{equation}
where\footnote{Notice that, if $X_{\mu \nu}$ is not symmetric, the expression of $Y_{\mu \nu}$ would need to be symmetrized in $\mu$ and $\nu$.}
\begin{equation}
    Y_{\mu \nu}
    \equiv
    [\mathcal{K}] \, g_{\mu \nu}
    -\mathcal{K}_{\mu \nu}
    + \left[ 
    (\mathcal{K}^2)_{\mu \nu}-[\mathcal{K}]\mathcal{K}_{\mu \nu}-\frac{1}{2}\mathcal{U}_2 \, g_{\mu \nu}
    \right]
    \ ,
\end{equation}
so that $T_{\mu\nu}^{\phi} \equiv  m^2 \MP^2 Y_{\mu \nu} / 2$ can be interpreted as the stress-energy tensor of the Stueckelberg sector, while $T_{\mu \nu}$ is the stress-energy tensor associated with ordinary matter content. Diffeomorphism invariance implies the covariant conservation for the Stueckelberg stress-energy tensor, $\nabla_{\mu}Y^{\mu}_{\nu} = 0$.~\footnote{
	Note that this equation is automatically satisfied when the Stueckelberg fields satisfy their Euler-Lagrange equations $\delta S / \delta \varphi^a = 0$,
	In particular, one can show that $\partial_{\nu}\phi^a \delta S / \delta \varphi^a = \nabla_{\mu}Y^{\mu}_{\nu}$.
	Therefore, covariant conservation of $Y^{\mu}_{\nu}$ is actually implied by the Euler-Lagrange equations, and that the two equations are actually equivalent.
}
As anticipated in Sec.~\ref{SecPX}, the imposition of the symmetry constraints on $(X^2)^\mu_\nu$ yields an effective perfect fluid stress-energy tensor for the Stueckelberg sector. More explicitly, it is easy to verify that the non-vanishing components of the $Y^{\mu}_{\nu}$ tensor are given by
\begin{equation}
        Y^0_0
        =
        3\left(2-3G+ G^2\right) \, \quad 
        Y^i_j
        =
        \left(6-3F-6G+ G^2+2FG\right)\delta_{ij}
        \ .
\end{equation}
Thus, we can identify the energy density and pressure of the Stueckelberg fluid as
\begin{equation}
    \begin{aligned}
        &
        \rho_{\phi}
        =
        -\frac{3m^2 \MP^2}{2}\left( 2-3G+ G^2\right) \ , \\
        &
        p_{\phi}
        =
        \frac{m^2 \MP^2}{2}\left( 6-6G+ G^2 - 3 F+2 FG\right)
        \ .
    \end{aligned}
\end{equation}
It is evident that both the energy density and pressure of the Stueckelberg sector are functions of $F(t)$ and $G(t)$ and therefore are homogeneous functions. 
In particular, $\rho_{\phi}$ only depends on $G(t)$. This fact is crucial to ensure the existence of a class of ``massive cosmologies" which can be fully determined without relying on the details of the choice of $\phi^a$ and $f_{ab}$.

Making explicit the components of the Einstein tensor and of the stress-energy tensor of the extra matter content (defined in terms of its density $\rho$ and pressure $p$), the Friedmann equations acquire the form:
\begin{equation}\label{eq:Friedman_minimal}
    \begin{aligned}
        & H^2
        =
        - \frac{m^2}{2}\left(2-3G+ G^2\right)+ \frac{\rho}{3\MP^2} 		\ , \\
        & 
        \frac{\ddot{a}}{a}
        =
        -\frac{m^2}{4}\left(4-3F-3G+2FG\right)-\frac{ 3p+ \rho}{6\MP^2}  \ . 
    \end{aligned}
\end{equation}
The continuity equation of the Stueckelberg sector can be written instead as
\begin{equation}\label{eq:minimalEoM}
    (3-2 G)
    \left[ 
    \dot G
    +H(G - F)
    \right] = 0
    \ .
\end{equation}
As already explained, this set of equations does not depend on the explicit form of the Stueckelberg fields or the choice of the reference metric. Interestingly,
we can classify the possible cosmologies according to which of the two factors in Eq.~\eqref{eq:minimalEoM} is imposed to vanish. As a result, we can identify two possible branches:
\begin{itemize}
    \item $\Lambda$-Branch : $G = 3 / 2$
    \ .
    \item Mixed Branch: $\dot G + H (G - F) = 0 \ $. 
\end{itemize}
We are now going to study these two branches separately.

\subsubsection{$\Lambda$-Branch}
This class of solutions is quite peculiar, since the specific value that $G$ takes 
cancels precisely any possible $F$ dependence in the pressure of the Stueckelberg fluid. As a consequence, $\rho_{\phi}$ and $p_{\phi}$ happen to be related by a barotropic equation of state with $w=-1$. 
This means that the stress-energy tensor of the Stueckelberg fluid is given by
\begin{equation}
	T^{\phi}_{\mu\nu}
	=
    \frac{m^2 \MP^2}{2}Y_{\mu\nu}
    =
    -
    \frac{3 m^2 \MP^2}{8} g_{\mu\nu}
    \ ,
\end{equation}
which corresponds to a cosmological constant of energy density
${3 m^2 \MP^2} / {8}$. 
As a result, the Friedmann equations acquire the form
\begin{equation}
    H^2 = \frac{m^2}{8}+ \frac{\rho}{3\MP^2} 
    \;, \quad
    \frac{\ddot{a}}{a} = \frac{m^2}{8}- \frac{3p+\rho}{6\MP^2}
    \ .
\end{equation}
It is interesting to notice that, once the residual matter content of the spacetime is specified, the cosmology is fully determined, in a way which seems to be insensitive to the details of the Stueckelberg configurations (as well as on the choice of the reference metric), provided we restrict only on families of configurations which satisfy the symmetry requirements\footnote{It is actually non-trivial to look for Stueckelberg profiles which fulfil homogeneity and isotropy of $(X^2)^{\mu}_{\nu}$ and, at the same time, support a dynamical cosmology.} \eqref{eq:constraintsstuck}. 
At the moment, we are assuming that there exists at least one choice of $f_{ab}$ and $\phi^a$ that realizes the initial symmetry conditions. 
Notice that, if more choices are possible, the background cosmology will not be able to distinguish between these different possibilities: the features of the $\Lambda$-Branch solution do not depend on the particular ansatz for $f_{ab}$ and $\phi^a$. As clarified in the introduction, the same self-accelerated solution was found by \cite{DAmico:2011eto}, even though the authors did start from a different ansatz for $X^{\mu}_{\nu}$\footnote{In \cite{DAmico:2011eto}, a spherically symmetric ansatz is made for the Stueckelberg fields and the reference metric is chosen to be Minkowski. As a consequence, $X^{\mu}_{\nu}$ is not diagonal and it does not enjoy the FLRW symmetries.}.

\subsubsection{Mixed Branch}

The second branch can accommodate different kinds of solutions.

\subsubsection*{Case $\dot{G}=0$}
The equation of motion forces either $F=G=\mathrm{const.}$ or $H=0$. In the second case, we hit again the No-Go theorem, while in the first scenario, the Stueckelberg fields behave again as a cosmological constant, whose energy density is
\begin{equation}
    \rho_{\phi}=-\frac{3m^2\MP^2}{2}(2-3G+G^2)
    \ .
\end{equation}

\subsubsection*{Case $\dot{G}\neq 0$}
Assuming now that $\dot G \neq 0$, we have that $G$ can be obtained in terms of $F$ and $a(t)$ by setting to zero the square bracket in Eq.~\eqref{eq:minimalEoM}. The resulting solution is
\begin{equation}\label{eq:solG_branchII}
    G(t) 
    =
    \frac{1}{a}
    \left(G_0+ \int_{t_0}^t F(t')\dot{a}(t') \de t'\right)
    \ ,
\end{equation}
where $G_0$ is a constant. 
In this case, more effort is needed to fully understand the cosmology of this family of solutions. 
Indeed, an extra relation between $F$ and $G$ is needed to express both $F$ and $G$ as functions of time and, subsequently, completely determine the evolution of the scale factor. Actually, this extra relation comes directly from the set of symmetry constraints \eqref{eq:constraintsstuck} and, in the specific case of Minkowski reference metric $f_{ab} = \eta_{ab}$, it takes the form (see App.~\ref{appconfig} for the derivation of the following relation)
\begin{equation}\label{eq:Minkrelation}
    a^2 G^2 
    =
    \lambda_1 \left(  \int_{t_0}^t F(t') \de t' + \sigma \right)^2 
    \ 
    \Longrightarrow
    \
    G
    =
    \pm
    \frac{\sqrt{\lambda_1}}{a} \, \bigg|  \int_{t_0}^t F(t') \de t' + \sigma \bigg|
    \;,
\end{equation}
where $\lambda_1$ and $\sigma$ are parameters that label the family of field configurations $\phi^a$ that satisfies the symmetry constraints. 

Combining Eq.~\eqref{eq:Minkrelation} with the equation of motion Eq.~\eqref{eq:solG_branchII} 
yields the following constraint
\begin{equation}
	\pm
\frac{\sqrt{\lambda_1}}{a}
	\ s \
	F
	=
	H F 
	\; , \quad \quad
	s \equiv {\rm sgn} \left[ \int_{t_0}^t F(t') \de t'  + \sigma \right]
	\ .
\end{equation}

The first possibility for solving the equation above is to set $F = 0$ and therefore have 
$G = g / a$, with $g \equiv \pm {\abs{\sigma}\sqrt{\lambda_1}}$
\footnote{It is assumed $\sigma\neq 0$, otherwise the Stueckelberg energy density would be negative.}. In this case, the behaviour of the Stueckelberg fluid is characterized by the following energy density and pressure
\begin{equation}\label{eq:rho_p_minimal_brII}
\begin{aligned}
    \rho_{\phi}
    &= 
    -\frac{3m^2 \MP^2}{2}
    \left(
    2
    -\frac{ 3 g}{a}
    + \frac{g^2}{ a^2}
    \right) 
    \ , \\ 
    p_{\phi}
    &=
    \frac{m^2\MP^2}{2}
    \left(
    6 
   - \frac{6 g}{a}
    + \frac{g^2}{ a^2}
    \right)
    \ .
\end{aligned}
\end{equation}
Notice, first of all, that the sign of $\rho_{\phi}$ is crucially dependent on the sign of G. Indeed, if $G<0$, the energy density of the Stueckelberg fluid is forced to be negative for the whole history of the spacetime. Moreover, the Stueckelberg sector behaves like a mixture of perfect fluids whose equations of state are respectively $w=-1,-{2} / {3},-{1} / {3}$. This means that at early times the Stueckelberg sector mainly contributes in the form of curvature, while the asymptotic behaviour at late times is the one of a negative cosmological constant, which is independent on $\sigma$ and $\lambda_1$\footnote{On phenomenological grounds, to have a curvature component $|\Omega_k| < 0.01$ today, one needs the combination $g^2 = \sigma^2 \lambda_1 \lesssim 0.01$. A stronger bound on $g$ could be obtained by imposing the energy component scaling as $1/a$ in Eq.~\eqref{eq:rho_p_minimal_brII} to be below today's observational limit.}.

Generally, such a rich scenario ends up being potentially pathological, because the energy density of the Stueckelberg sector sees its sign varying throughout the history of the universe. Indeed, the energy density is positive only in the interval $g / 2 < a < g^2$.
It will be shown that, in the context of the full theory, an additional perfect fluid component (with an equation of state $w=0$) appears in the Stueckelberg sector. 

In addition to the two previous scenarios, the Mix-Branch can accommodate for another class of solutions, which is identified by $F\neq 0$. In this case, Eq.~\eqref{eq:Minkrelation} transforms into a constraint on the Hubble parameter:
\begin{equation}\label{eq:brII_solH}
    H = \pm \frac{\sqrt{\lambda_1}}{a} s \ .
\end{equation}
Note that, demanding $H > 0$ would impose that $\pm s = 1$.
From Eq.~\eqref{eq:brII_solH} we see that, one more time, the Stueckelberg sector imposes a constraint on the cosmology. A No-Go theorem knocks again at the door, forcing the second derivative of the scale factor to vanish, $\ddot a  = 0$, and leading to a universe that expands/contracts in a uniform way (as $a \propto t$).
In this particular situation, the Friedmann equations in Eq.~\eqref{eq:Friedman_minimal} become a set of algebraic equations for $F$ and $G$.

If other matter components were present, both $F$ and $G$ would depend on the rest of the matter content of the spacetime. This behaviour is crucial to enable the Stueckelberg fluid to cancel the influence of the other fluids and impose the No-Go constraint on $\ddot{a}$.\\
It is worth stressing that the features of this Branch, when $\dot{G}\neq 0$, strongly rely on the choice of the reference metric. A different ansatz for $f_{ab}$ would, in principle, lead to a different extra relation between $F$ and $G$, resulting in a different behaviour of the Stueckelberg fluid. Notice that similar branches were obtained in the context of bi-gravity \cite{Comelli_2012}, with an FLRW ansatz for both metric tensors.

\begin{table}[t]
    \centering
   \begin{tabular}{ |c|c|c|c|c|c| } 
	\hline
	\rowcolor{Gainsboro!60}
	Branches & $G$   & $F$ &  $f_{ab}$ & Cosmology 							\\
	\thickhline
	$\Lambda$-Branch		 & $3/2$ & $F(G,f_{ab})$   &     $-$     &  $\Lambda$			\\
	\hline
	\multirow{4}{*}{\centering Mixed} 
	& $\propto 1/a$ &     0      & $\eta_{ab}$ &    Mix 						\\ 
	& $\propto 1/a$ & $\neq 0$  & $\eta_{ab}$ & $\ddot{a}=0$				\\
	& $\mathrm{const}$ & $G$ &  $-$  &   $\Lambda$ 					\\ 
	& $\mathrm{const}$ & $\neq G$& $-$&$H=0$						\\ 
	\hline
\end{tabular}
    \caption{~Different branches and their cosmological behaviour. The $\Lambda$-Branch is the self-accelerated background solution, in which the Stueckelberg sector behaves as a cosmological constant. The Mixed Branch can accommodate different classes of solutions. Among them, a dynamical background in which the Stueckelberg fields behave as a mixture of perfect fluids is obtained once the choice $f_{ab}=\eta_{ab}$ is specified. Similarly, also the $\ddot{a}=0$ scenario requires a specification of the reference metric. Indeed, once $f_{ab}$ is specified, an extra relation between $F$ and $G$ is provided by the symmetry conditions imposed on $X^{\mu}_{\nu}$, enabling to determine $F$ in terms of $G$.
    All the other scenarios do not rely on a specific ansatz for $f_{ab}$ and $\phi^a$ (provided they satisfy the consistency relations \eqref{eq:constraintsstuck}). 
   }
    \label{tab:my_label}

\end{table}

\subsection{dRGT: Full Theory}\label{fullModel}
It is now time to generalise the previous results to the full dRGT model, in which both the cubic and quartic potentials $\mathcal{U}_3$ and $\mathcal{U}_4$, defined in Eq.~\eqref{eq:dRGT_potentials}, are included.
As before, the starting point is the imposition of the symmetry conditions \eqref{eq:sqroot} on $X^{\mu}_{\nu}$. It is not difficult to check that the explicit expressions of the potentials in terms of the functions $F$ and $G$ which encapsulate the net effect of the inhomogeneous Stueckelberg configurations are
\begin{equation}\label{eq:pot_full_th_expr}
    \begin{aligned}
		&\mathcal{U}_2 = -6 (G-1) (F+G-2)
		\ , \\
		&\mathcal{U}_3 = 6 (G-1)^2 (3 F+G-4)
		\ , \\
		&\mathcal{U}_4 = -24 (F-1) (G-1)^3
		\ .
    \end{aligned}
\end{equation}
The contribution of the additional potentials to the Stueckelberg stress-energy tensor can be straightforwardly evaluated\footnote{Notice that $X_{\mu \nu}=X_{\nu\mu}$ as a consequence of our initial assumption about $X^{\mu}_{\nu}$ being FLRW-compatible.}, yielding 
\begin{equation}
\begin{aligned}
    &Y_{\mu \nu}=\alpha_2 \left[  \mathcal{K}^2_{\mu \nu}-\mathcal{K}_{\mu \nu}+(g_{\mu \nu}-\mathcal{K}_{\mu \nu})[\mathcal{K}]- \frac{1}{2}\mathcal{U}_2g_{\mu \nu} \right]+\\
    &+\alpha_3\left[- 3(\mathcal{K}_{\mu}^{\rho}- [\mathcal{K}]\delta^{\rho}_{\mu})(\mathcal{K}_{\rho \nu}^2-\mathcal{K}_{\rho \nu})- \frac{3}{2}\mathcal{U}_2 (g_{\mu \nu}- \mathcal{K}_{\mu \nu})- \frac{1}{2}\mathcal{U}_3 g_{\mu \nu}\right]+\\
    &+\alpha_4\left[6(2\mathcal{K}_{\mu}^{\sigma}\mathcal{K}_{\sigma}^{\rho}-2[\mathcal{K}]\mathcal{K}^{\rho}_{\mu}- \mathcal{U}_2 \delta^{\rho}_{\mu})(\mathcal{K}_{\rho \nu}^2-\mathcal{K}_{\rho \nu})-2 \mathcal{U}_3 (g_{\mu \nu}-\mathcal{K}_{\mu \nu}) -\frac{1}{2}\mathcal{U}_4g_{\mu \nu}\right]
    \ .
    \end{aligned}
\end{equation}
Given the symmetries we are imposing on $X^{\mu}_{\nu}$, the Stueckelberg sector effectively behaves as a perfect fluid also in the full theory. Making use of the explicit expressions for the potentials Eq.~\eqref{eq:pot_full_th_expr} and of Eq.~\eqref{eq:sqroot}, we obtain the following values for the non-vanishing components of the $Y_{\mu \nu}$ tensor
\begin{equation}\label{eq:Y_full_model}
    \begin{aligned}
        &
        Y_{00}
        =
        \left(-6 \alpha _2-12 \alpha _3-12 \alpha _4\right)
        +\left(-3 \alpha _2-18 \alpha _3-36 \alpha _4\right) G^2
        +\left(9 \alpha _2+27 \alpha _3+36 \alpha _4\right) G+\left(3 \alpha _3+12 \alpha _4\right) G^3
        \ , \\
		&
		Y_{ij}
		=
		\delta_{ij}a^2 
		\bigg[
		\left(6 \alpha _2+12 \alpha _3+12 \alpha _4\right)
		+\left(\alpha _2+6 \alpha _3+12 \alpha _4\right) G^2
		+\left(-6 \alpha _2-18 \alpha _3-24 \alpha _4\right) G
		+ 
		\\
		&
		+F \bigg( 
		\left(-3 \alpha _2-9 \alpha _3-12 \alpha _4\right)
		+\left(-3 \alpha _3-12 \alpha _4\right) G^2
		+\left(2 \alpha _2+12 \alpha _3+24 \alpha _4\right) G
		\bigg)
		\bigg]
		\ .
    \end{aligned}
\end{equation}
These expressions can be mapped to the energy density and the pressure of the Stueckelberg fluid as
\begin{equation}
    	\rho_{\phi} = \frac{\MP^2 m^2}{2} Y_{00} 
    	\ , \quad 
        p_{\phi} = \frac{1}{3}\frac{\MP^2m^2}{2}Y^i_i
        \ .
\end{equation}
Lastly, the continuity equation for the Stueckelberg fields leads to the following relation:
\begin{equation}\label{eq:continuity_full_th}
    \begin{aligned}
       \Big[
        3 \left(\alpha _3+4 \alpha _4\right) G^2
        -2 \left(\alpha _2+6 \alpha _3+12 \alpha _4\right) G
        +3 \left(\alpha _2+3 \alpha _3+4 \alpha _4\right)
       \Big]
        \Big[
        \dot{G}
        + H (G-F)
        \Big]
        = 0
        \ . 
    \end{aligned}
\end{equation}
Notice that, similarly to the minimal model case, the continuity equation factorizes. This fact will allow us to identify two different cosmological branches, obtained by setting each of the two square brackets in Eq.~\eqref{eq:continuity_full_th} to zero.

Remarkably, the structure of the equation is very similar to the case of the minimal model Eq.~\eqref{eq:minimalEoM}. In particular, the second factor in Eq.~\eqref{eq:continuity_full_th} is identical to the one we already encountered while the first, once again, does not depend on the function $F$. 
It can be expected, then, that the qualitative features of the different cosmological branches will be extremely similar to those discussed in the previous section. 
Explicitly, the two branches are identified by the following conditions:
\begin{itemize}
    \item $\Lambda$-Branch: 
    $ \ 
    3 \left(\alpha _3+4 \alpha _4\right) G^2
    -2 \left(\alpha _2+6 \alpha _3+12 \alpha _4\right) G
    +3 \left(\alpha _2+3 \alpha _3+4 \alpha _4\right)
    =
    0 \ $.
    \item Mix Branch: $ \ \dot G + H(G-F) = 0 \ $. 
\end{itemize}
Their properties are summarized in Tab.~\ref{tab:my_label}.

\subsubsection{$\Lambda$-Branch}
\begin{figure}[t!]
    \centering
    \includegraphics[width=\textwidth]{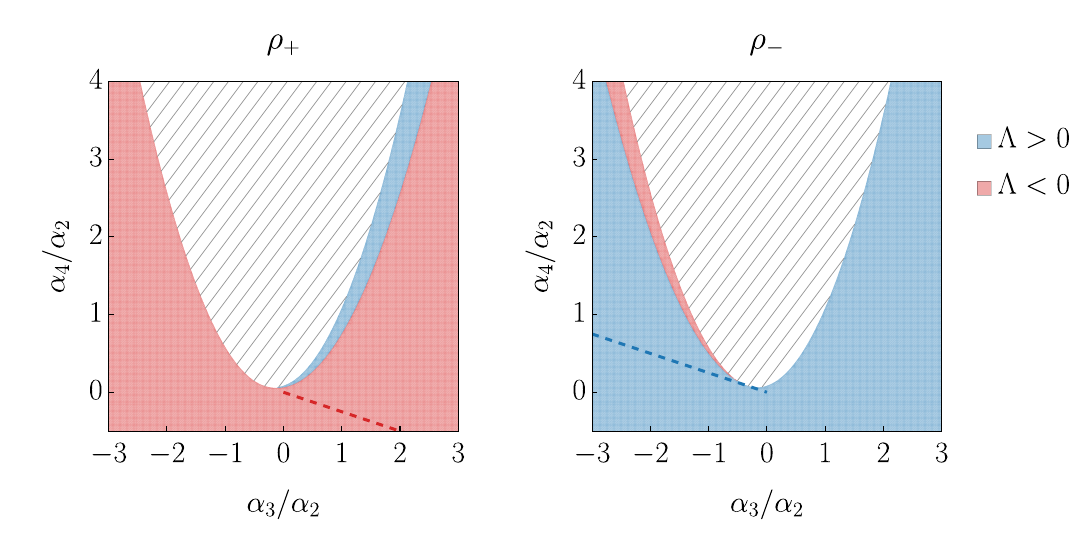}
    \caption{~Regions of parameter space for which the Stueckelberg sector behaves as a positive (blue) or negative (red) cosmological constant in the $\Lambda$-Branch. The dashed line corresponds to the condition $\alpha_3+4\alpha_4=0$, for which the equation that identifies branch I becomes linear in $G$. In this special case, the Stueckelberg sector behaves as a positive cosmological constant only when $\alpha_3>-1/3$. The white hatched region corresponds to the values of parameters for which the discriminant $\Delta$ is negative (and thus yields complex solutions $G_{\pm}$).
    Along the dashed lines $\rho_{\phi}$ diverges, with a sign indicated by the color of the line.}
    \label{Plotbranch1}
\end{figure}

As before, the first branch is identified by the vanishing condition of the above algebraic equation for $G$. This is nothing but a quadratic equation for $G$, and its roots are simply given by
\begin{equation}\label{rootG}
    G_{\pm}
    =
    \frac{
    \alpha _2+6 \alpha _3+12 \alpha _4
    \pm 
    \Delta
    }
    {3 \left(\alpha _3+4 \alpha _4\right)}
    \ , 
    \quad 
    \Delta 
    \equiv 
    \sqrt{\alpha _2^2+3 \left(\alpha _3-4 \alpha _4\right) \alpha _2+9 \alpha _3^2}
    \ .
\end{equation}
It is important to point out that, in general, not every region of the parameter space of the theory will allow for the existence of such roots. Indeed the above solutions will be real only when $\Delta > 0$.
Similarly to the minimal-model case of the previous Section, once the above solutions for $G$ are plugged inside $Y_{\mu \nu}$, any $F$ dependence in the pressure term is completely washed away, ultimately leading to the following expressions for $\rho_{\phi}$ and $p_{\phi}$:
\begin{equation}
\begin{aligned}
	&\rho_\phi 
	= \frac{\left(\alpha _2+3 \alpha _3 \pm\Delta \right) \left(\mp3 \alpha _3 \Delta -\alpha _2 \left(3 \alpha _3-36 \alpha _4\pm\Delta \right)-2 \alpha _2^2-18 \alpha _3^2+\Delta ^2\right)}{9 \left(\alpha _3+4 \alpha _4\right){}^2} \  , \\
	&
	p_{\phi} = -\rho_{\phi}
	\ .
\end{aligned}
\end{equation}
As explicitly shown by the above expression, the value of the cosmological constant now depends on the parameters of the theory. In particular, we can write $\Lambda = \rho_\phi / \MP^{2}$,
and explore the parameter space region for which $\Lambda > 0$ (see Fig.~\ref{Plotbranch1}).

As before, the cosmology of the $\Lambda$-Branch is fully determined by the Stueckelberg equation of motion. On the same lines as the discussion of the minimal model, therefore, this branch needs no extra relation between $F$ and $G$ to obtain the evolution of the scale factor. 
In particular, neither $f_{ab}$ nor $\phi^a$ need to be specified. 
Compatibility with the symmetry conditions of Eq.~\eqref{eq:constraintsstuck} will however provide a relation between $F$ and $G$ once the reference metric is specified.
\subsubsection{Mixed Branch}
Also in the full theory, this second branch is identified by the condition $\dot G + H(G-F) = 0$.
As already discussed in the context of the minimal model, an extra relation between $F$ and $G$ is required to express both the functions in terms of the scale factor (or, equivalently, in terms of $t$) and to fully determine the cosmological history of spacetime. 
Again, we fix the reference metric to be Minkowski spacetime. As a consequence, the extra relation between $F$ and $G$ takes the form Eq.~\eqref{eq:Minkrelation} and it can be combined with the continuity equation for the Stueckelberg stress-energy tensor to determine the time evolution of the functions $F$ and $G$. 

Following the steps of Sec.~\ref{sub:Minimal} we obtain a very similar set of possible cosmological scenarios. Among them, the class of solutions identified by $F\neq 0$ yields a cosmology in which the Hubble parameter is constrained to be $H\propto 1 / a$. As a consequence, the Stueckelberg sector imposes a vanishing condition on the second derivative of the scale factor, giving rise to a linear expansion expansion/contraction. 
Solving the Friedmann equations enables to obtain the explicit expression of $F$ and $G$ in terms of the scale factor and of the energy density and pressure of matter (if present). As already pointed out, in this peculiar scenario, the Stueckelberg fluid counterbalances the effect of the other perfect fluids to impose the $\ddot{a}=0$ constraint.

The other, more interesting, cosmological scenario corresponds to the situation
\begin{equation}
    F = 0 \ , 
    \quad  
    G = \frac{g}{a}
    \ ,
\end{equation}
where the constant $g \equiv \pm |\sigma| \sqrt{\lambda_1}$ can take either sign.
As before, for this class of solutions, the Stueckelberg sector consists of a mixture of perfect fluids with different equations of state. 
Using the above expression for $G$ inside Eq.~\eqref{eq:Y_full_model},
it is straightforward to see that the energy density and pressure of the Stueckelberg fluid are given by
\begin{equation}
    \begin{aligned}
    \rho_{\phi}
    &=
    \frac{3 m^2 \MP^2}{2}
    \bigg[
    -2(\alpha_2+2\alpha_3+2\alpha_4)
    +\frac{3 g}{a} (\alpha_2+3 \alpha_3+4 \alpha_4)  
    - \frac{g^2}{a^2} (\alpha_2+6 \alpha_3+12\alpha_4) 
    +
    \\
    &
    \hspace{1.7cm}
    + \frac{g^3}{a^3}(\alpha_3+4\alpha_4)
    \bigg]
    \ ,
    \\
    p_{\phi}
    &= 
    \frac{m^2\MP^2}{2}
    \bigg[ 
    6 (\alpha_2+2 \alpha_3+2 \alpha_4)
    - \frac{6 g}{a} (\alpha_2+3 \alpha_3+4 \alpha_4) 
    +\frac{g^2}{a^2} (\alpha_2+6 \alpha_3+12 \alpha_4)  
    \bigg]
    \ .
    \end{aligned}
\end{equation}
It is evident that the rich mixture of fluids in the Stueckelberg sector consists of four components, with equations of state $w = -1$, $-2/3$, $-1/3$ and $0$.
The presence of non-relativistic matter in the Stueckelberg fluid is a novelty of the full model. Indeed, the energy density of this component crucially depends on $\alpha_3$ and $\alpha_4$, and therefore it did not appear in the minimal model. This peculiar scenario thus sees the Stueckelberg fluid behaving at early times mainly as dust, making it possible (in principle) to ascribe to the new polarizations of the massive graviton a partial contribution to the Dark Matter content. At very late times, on the other hand, the Stueckelberg fluid eventually dominates the energy budget of the universe in the form of a cosmological constant with value
\begin{equation}
    \Lambda = -3m^2(\alpha_2+2 \alpha_3+2\alpha_4) \ .
\end{equation}
Contrary to what happened in the minimal model, there exists now a choice of parameters for which $\Lambda$ is positive. It is indeed sufficient to choose
\begin{equation}
    \alpha_3< \alpha_4+\frac{\alpha_2}{2}.
\end{equation}

\begin{figure}
    \includegraphics[width=\textwidth]{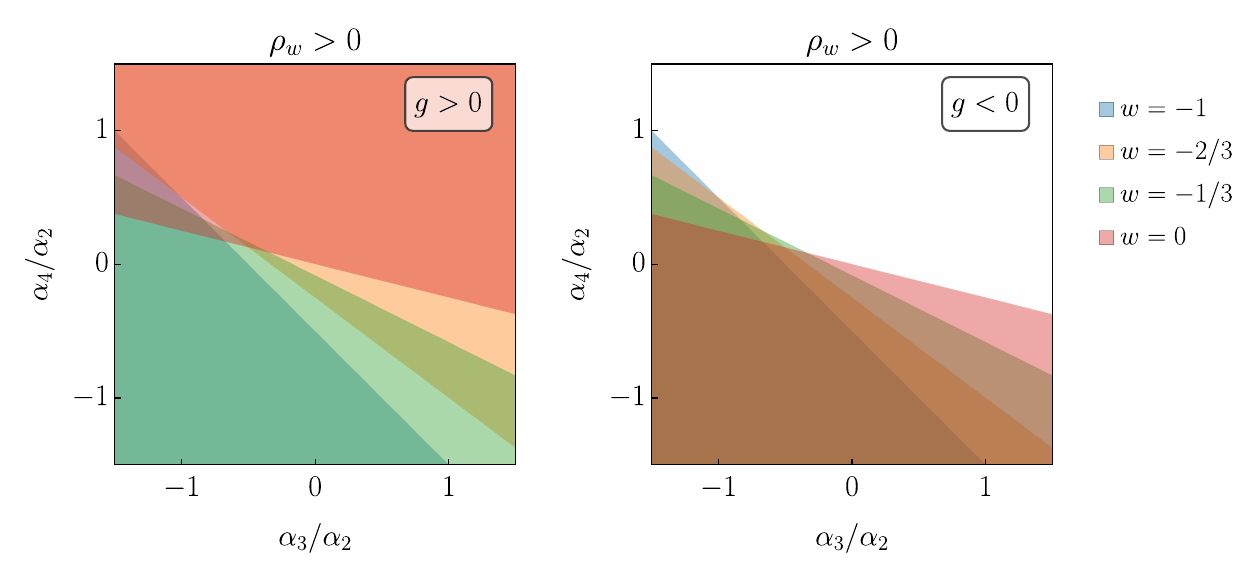}
    \caption{~Region of parameter space for which each individual $\rho_w$ is positive. For $g >0$ (left), no combination of parameters ensures the positivity of all the fluid components. On the contrary, when $g <0$ (right panel), there is a region of parameter space for which positivity of all the energy density components is fulfilled (the trapezoid resulting from the intersection of all the colored regions).}
    \label{Plotbranchmix}
\end{figure}

Notice that demanding the cosmological constant to be due entirely by the Stueckelbergs fixes $m$ to be proportional to $H_0 \sim 10^{-33} \ {\rm eV}$.
Moreover, in the presence of dust and radiation, additional bounds on $g$ can be imposed by requiring the other components of $\rho_{\phi}$ not to contribute too much to curvature and to avoid a large $1/a$ density.

As mentioned in the introduction, this class of background solutions was obtained in \cite{Volkov:2012zb} under the assumption of a spherically symmetric profile for the Stueckelberg fields and the choice of a diagonal reference metric.

Up to now, we have explored the features of viable dynamical dRGT cosmologies at the background level. The next section will address the stability of these solutions against linear perturbations in the absence of extra matter fields\footnote{The inclusion of a generic fluid into the description has no impact on the qualitative behaviour of the metric perturbations, as discussed in App.~\ref{apppert}.}.

\section{Perturbations in the Unitary Gauge}\label{sec:perturbs}

We will tackle the perturbative analysis in the unitary gauge, where the Stueckelberg perturbations are set to zero and all the degrees of freedom live inside the metric.
To do so we identify the tensor $\Sigma_{\mu \nu} \equiv \partial_{\mu}\phi^a \partial_{\nu}\phi^b f_{ab}$ with its background counterpart $\bar{\Sigma}_{\mu \nu}$ and split the physical metric as $g_{\mu \nu}=\bar{g}_{\mu \nu}+h_{\mu \nu}$, where the perturbations $h_{\mu \nu}$ are SVT decomposed and parametrized by
\begin{equation}\label{eq:SVT}
    h_{00} = A \ , \quad 
    h_{0i} = \hat{B}_i + \partial_i B \ , \quad 
    h_{ij } = C \, \delta_{ij} 
    		+ \partial_{\langle i}\partial_{j\rangle} E 
    		+\partial_{(i}\hat{E}_{j)}+\hat{h}_{ij} \ ,
\end{equation}
where $\hat{B}_i$ and $\hat{E}_i$ are transverse 3-vectors and $\hat{h}_{ij}$ is a transverse and traceless 3-tensor. Notice that Eq.~\eqref{eq:squared} translates into the requirements

\begin{equation}
    \bar{\Sigma}_{00} = -F^2 \ , \quad 
 	\bar{\Sigma}_{0i} = 0 \ ,    \quad 
	\delta^{ik} \bar{\Sigma}_{kj} = a^2 G^2 \delta^i_j
	\ . 
\end{equation}
Writing the inverse metric as $g^{\mu \nu}=\bar{g}^{\mu \nu}+H^{\mu \nu}$, where $H^{\mu \nu}=-\bar{g}^{\mu \alpha}h_{\alpha \beta}\bar{g}^{\beta \nu}+\bar{g}^{\mu \alpha}h_{\alpha \beta}\bar{g}^{\beta \gamma}h_{\gamma \delta}\bar{g}^{\delta \nu}+O(h^3)$, 
the explicit expression of $X^2$ is given by 
\begin{displaymath}
    \begin{aligned}
        &(X^2)^0_0=F^2-F^2H^{00} \ , \\
        &(X^2)^0_i=G^2 a^2H^{0j}\delta_{i j}=G^2 h_{0i} \ , \\
        &(X^2)^i_0=-F^2H^{i0} \ , \\
        &(X^2)^i_j=G^2 \delta^i_j+G^2 a^2H^{ik}\delta_{kj}=G^2 \delta^i_j- \frac{G^2}{a^2}\delta^{ik}h_{kj} \ .
    \end{aligned}
\end{displaymath}
To construct the square root tensor we parametrize its components as
\begin{displaymath}
X^{\mu}_{\nu}=
    \begin{bmatrix}
        \Omega & G_j\\
        H^i &K^i_j 
    \end{bmatrix}
    \ .
\end{displaymath}
Taking the square, we arrive at the system of equations
\begin{equation}
    \begin{cases}
        \Omega^2+G_jH^j=(X^2)^0_0\\
        G_j(\delta^j_i \Omega+K^j_i)=(X^2)^0_i\\
        (\delta^i_j \Omega+K^i_j)H^j=(X^2)^i_0\\
        H^i G_j +K^i_k K^k_j=(X^2)^i_j
    \end{cases},
\end{equation}
whose solution is
\begin{equation}
\begin{aligned}\label{eq:Xmunu}
    &\Omega
    =
    F\left[1+\frac{h_{00}}{2}+\frac{3}{8}h_{00}^2-\frac{F(F+2G)}{2(F+G)^2}\bar{g}^{ij}h_{0i}h_{0j}\right]
    \ , \\
    &
    G_i
    =
    \frac{G^2}{F+G}h_{0i}\>,\quad 
    H^i
    =
    -\frac{F^2}{F+G}\bar{g}^{ij}h_{j0}
    \ , 
    \\
    &
    K^i_j
    =
    G\left[\delta^i_j-\frac{1}{2}\bar{g}^{ik}h_{kj}+\frac{3}{8}\bar{g}^{ik}\bar{g}^{mn}h_{km}h_{nj}-\frac{G(G+2F)}{2(F+G)^2}\bar{g}^{ik}h_{k0}h_{0j}\right]
    \ . 
\end{aligned}
\end{equation}
Plugging the above expression inside \eqref{eq:dRGT_action} yields the quadratic Lagrangian for the perturbations
\begin{equation}
\mathcal{L}^{(2)}=\mathcal{L}^{(2)}_{\rm S}+\mathcal{L}^{(2)}_{\rm V}+\mathcal{L}^{(2)}_{\rm T} \ ,
\end{equation}
where we have distinguished the quadratic Lagrangians of the scalar ($\rm S$), vector ($\rm V$) and tensor ($\rm T$) degrees of freedom. We now focus on each individual Lagrangian and study the stability of the perturbations on the $\Lambda$ and Mixed branches. Notice that one might be tempted to argue that $X^{\mu}_{\nu}$ depends non-analytically on the metric perturbations on the Mixed-Branch, since the pull-back of the field-space metric $f_{ab}$ is degenerate on that background, and hence that perturbations are strongly-coupled on that background. However, Eq.~\eqref{eq:Xmunu} shows that the components of $X^{\mu}_{\nu}$ are analytic functions of the metric perturbations even on the Mixed Branch. As a consequence, a perturbative analysis is necessary in order to unveil potential pathologies of this background solution.
\subsection{Tensors}
The tensor Lagrangian is 
\begin{equation}\label{eq:L_2_tensor}
	\mathcal{L}^{(2)}_{\rm T}
	=
	\frac{1}{4a}
	\left[
	\dot{\hat{h}}_{ij}^2
	-\frac{1}{a^2}(\partial_k \hat{h}_{ij})^2
	-2(2H^2+\dot{H})\hat{h}_{ij}^2
	-2 m^2_{\hat{h}} \hat{h}_{ij}^2
	\right] 
	\ ,
\end{equation}
where we have defined the background-dependent mass
\begin{equation}
\begin{aligned}
     m^2_{\hat{h}}
     &=
     \frac{m^2}{4}
     \Big\{
     \alpha _2 \left[3 F (G-2)+G^2-9 G+12\right]
     -3 \alpha _3 \left[F \left(G^2-6 G+6\right)-2 G^2+9 G-8\right]
     -\\
     &-12 \alpha _4 (F-1) \left(G^2-3 G+2\right)
     \Big\}
     \ ,
\end{aligned}
\end{equation}
and contractions of spatial indexes are done with the Kronecker delta. To make sure the backgrounds are healthy we shall ask the perturbations to be free from ghost and gradient instabilities. The absence of such pathologies is evident from the kinetic term in Eq.~\eqref{eq:L_2_tensor}. Therefore, we do not obtain any constraint on the background quantities from tensor perturbations.

\subsection{Vectors}
In the case of vector modes, the requirement of having healthy perturbations puts strong constraints on the background as we are going to show.
The vector Lagrangian is given by
\begin{equation}
\begin{aligned}\label{eq:lag_2_vect_0}
	\mathcal{L}^{(2)}_{\rm V}
	&=
	\frac{1}{8a}
	\left[
	(\partial_k \dot{\hat{E}}_j)^2
	-2(2H^2+\dot{H})(\partial_k \hat{E}_j)^2
	-4 m^2_{\hat{E}_i}(\partial_k \hat{E}_i )^2
	\right]
	+
	\\
	&
	+\frac{1}{2a}(\partial_k \hat{B}_i)^2 
	+3aH^2\hat{B}_i^2
	-\frac{1}{2} a m^2_{\hat{B}_i}\hat{B}_i^2
	+\frac{1}{2a}
	\left(
	2H\partial_k \hat{B}_j\partial_k \hat{E}_j-\partial_k \hat{B}_j\partial_k \dot{\hat{E}}_j
	\right)
	\ ,
\end{aligned}
\end{equation}
with
\begin{equation}
    \begin{aligned}
        m^2_{\hat{B}_i}
        &=
        -\frac{m^2}{F+G}
        \Big\{
        \alpha _2 
        \left[
        3 F \left(G^2-3 G+2\right)
        +G \left(G^2-6 G+6\right)
        \right]
        -3 \alpha _3  (G-1) 
        \big[
        F \left(G^2-5 G+4\right)
        \\
        &
        -2 (G-2) G
        \big]
        -12 \alpha _4 (G-1)^2 [F (G-1)-G]
        \Big\} 
        \ ,
        \\
        m^2_{\hat{E}_i}
        & =\frac{m^2_{\hat{h}}}{2}
        \ . 
    \end{aligned}
\end{equation}
Notice that $\hat B_i$ has no time kinetic term and therefore represents a constrained variable, that can be integrated out in terms of $\hat E_i$.
Going to Fourier space, the equation of motion for $\hat B_i$ gives
\begin{equation}
    \hat{B}_i
    =
    -\frac{k^2 / a^2 \left(2 \hat{E}_i H- \dot{\hat{E}}_i\right)}
    {2 \left(6  H^2-m^2_{\hat{B}_i}+k^2 / a^2\right)}
    \ .
\end{equation}
By substituting this solution back into Eq.~\eqref{eq:lag_2_vect_0}, we obtain the quadratic Lagrangian of the dynamical vector:

\begin{equation}
\label{eq:vect_def_params}
	\mathcal{L}^{(2)}_{\hat{E}_i}
	=
	K_{\hat{E}_i}^2\dot{\hat{E}}_i^2-Q^2_{\hat{E}_i}\hat{E}_i^2
	\ ,
\end{equation}
where we have defined
\begin{equation}
\begin{aligned}
	&K_{\hat{E}_i}^2
	\equiv 
	-\frac{k^2}{2 a^2}\frac{a\dot{H}}{k^2 / a^2-4\dot{H}}
	\ ,
	\\
    &
    M^2_{\hat{E}_i}
    \equiv
    2 m^2_{E_i} a^2 (k^2 / a^2 - 4 \dot{H} )
    -4 a^2 \dot{H} \big(\dot{H}+2 H^2\big)
    +k^2 \big(\dot{H}+4 H^2\big) 
    \ , 
    \\
    &
    Q^2_{\hat{E}_i}
    =
    \frac{k^2}{4a^3(k^2 / a^2 - 4 \dot{H})}
    \left[
    M^2_{E_i}
    +k^2\dot{H}
    -k^2H
    \frac{
    \partial_t(a^3(k^2 / a^2 - 4 \dot{H}))
    }
    {a^3(k^2 / a^2-4 \dot{H})}
    \right]
    \ .
    \end{aligned}
\end{equation}
We will now study the stability of $\hat E_i$ for the different backgrounds obtained in Sec.~\ref{sec:dRGT_bkg}, the $\Lambda$ and Mixed Branch.
\subsubsection{$\Lambda-$Branch}
On the $\Lambda-$Branch background, one has $\dot H = 0$ and therefore the kinetic term of vectors is switched off ($K_{\hat{E}_i}^2 = 0$).
As a consequence, the vector mode is non-dynamical and, in absence of matter perturbations, its equation of motion leads to $\hat{E}_i = 0$. 
The absence of the vector degree of freedom at the quadratic level is a sign of strong coupling. Therefore we conclude that the $\Lambda$-Branch is not a viable background.

\subsubsection{Mixed Branch}
The Mixed Branch supports the propagation of the vector mode. Indeed, on this background $\dot{H}\neq 0$, which means that the kinetic term of $\hat{E}_i$ is non-zero.

The healthiness requirements of the vector sector are realized, first of all, by the absence of ghost instabilities in the sub-horizon limit. This is achieved if the kinetic term has a positive sign (i.e.~$K_{E_i}^2 > 0$) when expanded at large $k/a$. From Eq.~\eqref{eq:vect_def_params} we then obtain $K_{E_i}^2 \simeq - a \dot H / 2$, implying that $\dot H$ must be positive.
This is a constraint on the parameter space of the theory, and it does not depend on the wave vector $k$.

Additional constraints come from the absence of gradient instabilities at small scales. In the large-$k$ limit, the vector Lagrangian acquires the form
\begin{equation}
\begin{aligned}
	\mathcal{L}^{(2)}_{\hat{E}_i}
	&\simeq
	-\frac{1}{2}a\dot{H}
	\left[
	\dot{\hat{E}}_i^2
	- \frac{c_s^2 k^2}{a^2} \hat{E}_i^2
	- \tilde{m}^2_{\hat{E}_i} \hat{E}_i^2
	\right]
	\ ,
\end{aligned}
\end{equation}
where we defined the speed of sound $c_s$ and the mass $\tilde m_{\hat E_i}$ as
\begin{equation}
    c_s^2
    \equiv 
    -\frac{2m^2_{E_i}+3H^2+2\dot{H}}{2\dot{H}}
    \ , \quad
    \tilde{m}^2_{\hat{E}_i}
    \equiv 
    -\frac{2}{\dot{H}}(3H^2\dot{H}+\dot{H}^2+H\ddot{H})
    \ .
\end{equation}
Since at small scales the $k^2$ will dominate over the mass, the absence of gradient instabilities implies the constraint $c_s^2 > 0$, which translates into
\begin{equation}
	\frac{2m^2_{E_i}+3H^2+2\dot{H}}{2\dot{H}} < 0
	\ . 
\end{equation}

\subsection{Scalars}

The Lagrangian of the scalar sector is given by
\begin{equation}
    \begin{aligned}
		\mathcal{L}^{(2)}_{\rm S}
		&=
		\frac{1}{6 a}
       \left[
       	(\nabla^2\dot{E})^2
        +\frac{1}{3 a^2}(\partial_i \nabla^2 E)^2
       	-\big(
       	4H^2+2\dot{H} + 3 m^2_E
       	\big) (\nabla^2 E)^2
       \right]
        - 
		\\
       &
       -\frac{a^3}{4} \left( 9 H^2  + 2 m_A^2\right) A^2
		+\frac{a}{2} \big(6 H^2 - m_B^2\big) (\partial_i B)^2
		-
		\\
		&
		-\frac{3}{2a} 
		\left[
			\dot{C}^2
			-\frac{1}{3 a^2}(\partial_i C)^2
			+\frac{1}{2} ( H^2 + 2\dot{H} + \frac{2}{3}m_C^2) C^2
		\right]
	   -
       \\
       &
       -2Ha \ \partial_k A \partial_k B
       +
       \frac{1}{2a}
       \left[
       a^2 (3 H^2 - 2 m_{AC}^2) AC
       - 6 a^2 H \dot{C}A
       -2\partial_k A\partial_k C
       \right]
       +
       \\
       &
       -\frac{1}{3a}\nabla^2 A\nabla^2 E
       +\frac{2}{a}\left(\dot{C} \nabla^2 B - 2 H C \nabla^2 B\right)
       -\frac{2}{3a}\left(B\nabla^4\dot{E} - 2 HB\nabla^4E \right)
       +\frac{1}{3a^3}C\nabla^4E       
       \ ,
    \end{aligned}
\end{equation}
where 
\begin{equation}
\begin{aligned}
	&
	m^2_E = \frac{2}{3}m^2_{\hat{h}}
	\ , \quad
	m^2_B = m^2_{B_i}
	\ ,
	\\
	&
	m_C^2
	=
	\frac{3 m^2}{4} 
	\Big\{
	\alpha_2 \left(3 F+G^2-6\right)
	-3 \alpha_3 \left[F \left(G^2-3\right)-2 G^2+4\right]
	-12 \alpha_4 (F-1) \left(G^2-1\right)
	\Big\}
	\ ,
	\\
	&
	m^2_A
	=
	-\frac{3}{4}m^2
	(G-1)
	\Big[
	2 \alpha _2
	+
	4 \left(\alpha _3+\alpha _4\right)
	+\left(\alpha _3+4 \alpha _4\right) G^2
	-\left(\alpha _2+5 \alpha _3+8 \alpha _4\right) G 
	\Big]
	=
	-\frac{3}{2}
	H^2
	\ ,
	\\
	&
	m^2_{AC}
	=
		\frac{3}{2}m^2 
		\left( 
		6 \left(\alpha _2+2 \alpha _3+2 \alpha _4\right)
		+\left(\alpha _2+6 \alpha _3+12 \alpha _4\right) G^2
		-6 \left(\alpha _2+3 \alpha _3+4 \alpha _4\right) G
		\right)
		\ .
    \end{aligned}
\end{equation}
In dRGT massive gravity, one expects only one scalar degree of freedom. Consistently, we find that the variables $A$, $B$, and $C$ are indeed non-dynamical and can be integrated out in terms of $E$. Let us see this explicitly.
Going to Fourier space and integrating out both $B_k$ and $A_k$ (they appear without time derivatives in the Lagrangian) we obtain the solutions
\begin{equation}
    \begin{aligned}
    &
    B_k
    =
    \frac{
    k^2(3C_k+k^2E_k)+3k^2a^2H(2E_k H-\dot{E}_k)-9a^2\dot{H}C_k
    }
    {6a^2H(3a^2\dot{H}-k^2)}
    \ , 
    \\
    &
    A_k
    =
    \frac{
    H[k^4\dot{E}_k+3\dot{C}_k(k^2-3a^2\dot{H})]-(2H^2+\dot{H})[k^4E_k +3C_k(k^2-3a^2\dot{H})]
    }
    {3a^2H^2(3a^2\dot{H}-k^2)}
    \ .
    \end{aligned}
\end{equation}
After replacing them inside the Lagrangian we get
\begin{equation}\label{eq:lag_scal_CE}
        \mathcal{L}_{C,E}^{(2)}
        =
        K_E^2\dot{E}^2-Q^2_EE^2-\tilde{Q}^2_C C^2
        \ .
\end{equation}
where
\begin{equation}\label{eq:def_ms_scalar}
    \begin{aligned}
    &
    K^2_E
    \equiv 
    -\frac{k^4 \dot{H}}{2a (k^2 / a^2-3\dot{H})}
    \ ,
    \\
    &
    M^2_E
    \equiv
    12 a^4 H^4 (2 k^2 / a^2-3 \dot{H})
    +3 a^4 H^2 \left(
    3 \dot{H} k^2 / a^2 
    -9 m_E^2 a \dot{H}
    -6 \dot{H}^2
    +3 a m_E^2 k^2 / a^2
    +k^4 / a^4
    \right)
    +k^4 \dot{H}
    \ ,
    \\
    &
    Q^2_E
    \equiv 
    \frac{k^4 / a^4}{18a H^2(k^2 / a^2 - 3\dot{H})}
    \left[
    M^2_E
    +H\partial_t(k^2(k^2+6a^2H^2)) 
    -\frac{k^2  \left(6 a^2 H^2+k^2\right)}{a^3(k^2-3 a^2 \dot{H})}
    \partial_t(a^3H(k^2-3 a^2 \dot{H}))
    \right]
    \ .
    \end{aligned}
\end{equation}
We do not write the explicit expression for $Q_C^2$ since it is not needed in the following.
Although not explicit initially, the Lagrangian \eqref{eq:lag_scal_CE} makes it clear that the variable $C$ is also not propagating. 
Therefore, the only dynamical scalar 
is $E$, whilst $C$ can be set to zero in the absence of matter sources.
The absence of ghost instabilities is achieved by the positivity of the kinetic term $K^2_E$.
Following the same reasoning as for vector modes, using Eq.~\eqref{eq:def_ms_scalar} we obtain that for large $k / a$ this imposes $\dot H < 0$. 

Next, we are going to impose the absence of gradient instabilities in the two Branches.

\subsubsection{$\Lambda$-Branch}
Notice that $K^2_E\propto \dot{H}$. This means that on the $\Lambda$-Branch, the kinetic term of the scalar mode is shut down, yielding a Lagrangian of the form
\begin{equation}
     \mathcal{L}^{\Lambda-\mathrm{Branch}}_{C,E}
     =
     -Q^2_E E^2 - \tilde{Q}^2_C C^2 
     \ .
\end{equation}As a consequence, we face the same problem that plagued the vector sector: strong coupling inhibits the propagation of the scalar mode at the linear level.

\subsubsection{Mixed Branch}
On the Mixed Branch, on the other hand, the kinetic term is nonzero and, in the large-$k$ limit, the Lagrangian of the scalar degree of freedom becomes
\begin{equation}
\begin{aligned}
	\mathcal{L}^{(2)}_{E}
	&\simeq
	-\frac{k^2}{2}a\dot{H}
	\left[
	\dot{E}^2
	-\left(
	\frac{c_s^2}{a^2}k^2+\tilde{m}^2_{E}
	\right)
	E^2
	\right]
	\ ,
\end{aligned}
\end{equation}
where we have defined
\begin{equation}
    c_s^2
    \equiv 
    -\frac{3m^2_{E}aH+6H^3+7\dot{H}H+\ddot{H}}{3H\dot{H}}
    \ ,
    \quad
    \tilde{m}^2_E
    \equiv
    -\frac{6H^3\dot{H}+7H\dot{H}^2+2H^2\ddot{H}+2\dot{H}\ddot{H}}{H\dot{H}}
    \ .
\end{equation}
Along the same lines as the previous discussion, the absence of gradient instabilities on sub-horizon scales is achieved by the requirement:
\begin{equation}
	\frac{3m^2_{E}aH+6H^3+7\dot{H}H+\ddot{H}}{3H\dot{H}} < 0
	\ .
\end{equation}
Noticeably, however, on the Mixed Branch background, the above inequality is exactly saturated, with the LHS exactly zero. As a consequence, the scalar mode has zero speed of sound on sub-horizon scales. 
Moreover, once interactions are included, an effective speed of propagation will be generated,
signalling the presence of strong coupling, as discussed in \cite{DAmico:2012qbv}.
In conclusion, also the Mixed Branch becomes untenable due to the strong coupling of perturbations.

\section{Conclusions}

This paper extended the approach proposed in \cite{Volkov:2012zb,Y_lmaz_2014_1} for the classification of the admissible flat FLRW cosmologies of dRGT massive gravity. Instead of specifying a particular ansatz for the profiles of the Stueckelberg fields and for the reference metric, FLRW symmetries are directly imposed at the level of the building block $X^{\mu}_{\nu}$ of the theory. 
As a consequence, the Stueckelberg sector effectively behaves as a perfect fluid whose energy density and pressure only depend on two homogeneous functions $F$ and $G$. 
The symmetry requirements imposed on $X^{\mu}_{\nu}$ result in a set of consistency constraints for the Stueckelberg and the reference metric. Remarkably, the Stueckelberg fields need to be inhomogeneous and/or anisotropic to satisfy these consistency constraints and sustain a dynamical flat FLRW background.

Assuming there exists at least a choice of $f_{ab}$ (and an associated choice of $\phi^a$) which satisfies these symmetry conditions,  different possible cosmological scenarios do emerge: 

\begin{enumerate}
\item The self-accelerating background solution, in which the Stueckelberg sector behaves as a CC, and for which we did not provide a simple configuration of $\phi^a$ when the reference metric is chosen to be Minkowski.

\item A scenario (derived under the assumption of $f_{ab}=\eta_{ab}$) in which the Stueckelberg sector behaves as a mixture of perfect fluids, for which an example of a viable choice of $\phi^a$ is provided in \cite{Volkov:2012zb}. 
\end{enumerate}

We also studied the behaviour of linear perturbations on top of these backgrounds, both in the presence and in the absence of matter. On the self-accelerated solution, the scalar and vector modes are strongly coupled, and only the tensor degrees of freedom propagate. The other background solution can accommodate propagation of the vector modes, while the scalar modes do not propagate on small scales (their speed of sound vanishes in the large-$k$ limit), signalling again strong coupling. Including matter into the description does not make any qualitative change, as all the metric perturbations will exhibit the very same behaviour.

Future work will address the possibility of extending the methodology presented in this paper, obtaining novel solutions by imposing the FLRW symmetries directly at the level of the Stueckelberg stress-energy tensor rather than on $X^{\mu}_{\nu}$. Unveiling novel cosmological solutions, these approaches may lead to interesting implications for dark energy, large-scale structure and cosmic tensions.

\section*{Acknowledgements}
We are very grateful to S.~Ansoldi, P.~Creminelli and M.~Delladio for the useful discussions.

\appendix 
\section{Viable Configurations}\label{appconfig}
The goal of this appendix is to show how to characterize the families of configurations of Stueckelberg fields that yield an $X^{\mu}_{\nu}$ tensor which satisfies the FLRW symmetries. We will focus on the choice $f_{ab}=\eta_{ab}$, but a similar construction can be carried out for any choice of the reference metric. As a starting point, we write \eqref{eq:squared} more explicitly in terms of the Stueckelberg fields $\phi^a$
\begin{equation}
    \begin{cases}
        F^2
        =
        (\dot{\phi}^0)^2 - \dot{\phi}^n\dot{\phi}^n \ ,
        \\
        \dot{\phi}^0\partial_i \phi^0
        =
        \dot{\phi}^n \partial_i \phi^n \ ,
        \\
        a^2 G^2\delta_{ij}
        =
        -\partial_i \phi^0 \partial_j \phi^0+ \partial_i \phi^n \partial_j\phi^n \ .
    \end{cases}
\end{equation}
Repeated spatial indices indicate contractions with the Kronecker delta.
Taking the time derivative of the second equation and the spatial derivative of the first one, we arrive at
\begin{equation}
    (\ddot{\phi}^n \dot{\phi}^0 - \ddot{\phi}^0 \dot{\phi}^n)
    \frac{\partial_i \phi^n }{(\dot{\phi}^0)^2}
    =0
    \ .
\end{equation}
Assuming that $\mathrm{Det}[\partial_i \phi^n]\neq 0$ enables us to simplify the above equation down to
\begin{equation}\label{Stuckrelations}
    \ddot{\phi}^n 
    = 
    \dot{\phi}^n\frac{\ddot{\phi}^0 }{\dot{\phi}^0}
    \quad
    \Longrightarrow
    \quad
    \dot{\phi}^n 
    =
    A^n(\vect x)\dot{\phi}^0 
    \ .
\end{equation}
Consequently, the time derivative of the spatial Stueckelbergs is proportional to the time derivative of the zeroth Stueckelberg via three functions of the spatial coordinates only, $A^n(\vect x)$. Plugging this information back into the above system and suppressing the explicit $x$-dependence, we get
\begin{equation}\label{setagain}
    \begin{cases}
        F^2
        =
        (\dot{\phi}^0)^2(1 - \vect A^2) 
        \ , \\
        \dot{\phi^0}(\partial_i \phi^0-A^n \partial_i \phi^n)
        =
        0
        \ , \\
        a^2 G^2\delta_{ij}
        =
        -\partial_i \phi^0 \partial_j \phi^0+ \partial_i \phi^n \partial_j\phi^n
        \ .
    \end{cases}
\end{equation}
Here and in the following $\vect A^2 \equiv A^{n} A^{n}$.
We will now investigate which families of profiles can give rise to the Mixed-Branch cosmology. 

\subsection{\texorpdfstring{$\Lambda$}{Lambda}-Branch}
As a first step, one can integrate Eqs.~\eqref{Stuckrelations} and \eqref{setagain} in time to fully express $\phi^n$ in terms of $\phi^0$:
\begin{equation}
    \phi^0
    =
    \frac{\sigma_u}{\sqrt{1-\vect{A}^2}}\int_{t_0}^t \de {t'} \, F({t'})
    + C(\vect x)
    \quad 
    \Longrightarrow
    \quad  
    \phi^n
    = 
    \sigma_u
    \frac{A^n(\vect x)}{\sqrt{1-\vect{A}^2}}
    \int_{t_0}^t \de {t'} \, F({t'})
    +B^n(\vect x)
    \ ,
\end{equation}
where $B^n(\vect x)$ and $C(\vect x)$ are functions to be determined and $\sigma_u$ is a sign.
Notice that the second equation of the set \eqref{setagain} imposes a relation between $A^n$, $B^n$ and $C$:
\begin{equation}
      \partial_i C = \vect{A}\cdot \partial_i \vect{B}
      \ ,
\end{equation}
where the scalar product is defined with respect to the Kronecker delta $\delta_{ij}$. Plugging the previous expressions inside the last equation of the symmetry constraints yields
\begin{equation}
\begin{aligned}
    a^2 G^2\delta_{ij}
    &=
    \partial_i \vect{B}\cdot \partial_j \vect{B}
    -(\vect{A}\cdot \partial_i \vect{B})(\vect{A}\cdot \partial_j \vect{B})
    +\left[ 
    \frac{
    \partial_i \vect{A}\cdot \partial_j \vect{B}
    +\partial_j \vect{A}\cdot \partial_i \vect{B}
    }
    {\sqrt{1-\vect A^2}}
    \right]
    \sigma_u \int_{t_0}^t \de t' \ F(t')
    +
    \\
    &
    + \left[ 
    \frac{\partial_i \vect{A}\cdot \partial_j \vect{A}}{1 - \vect A^2}
    +\frac{\partial_i \vect A^2 \partial_j \vect A^2 }{4(1 - \vect A^2)^2}
    \right]
    \left(\int_{t_0}^t \de t' \ F(t')\right)^2
    \ .
    \end{aligned}
\end{equation}
At this stage, we shall notice that the LHS is a homogeneous function of time, while the RHS is a sum of three factorizable contributions, with the time dependence completely stored inside the factors $(\int \de t' \ F(t'))^n$. As a consequence, the only possible way in which the above equation can be satisfied is to impose the following conditions, for some constants $\lambda_1$, $\lambda_2$ and $\lambda_3$:
\begin{equation}\label{eq:sys_appA}
    \begin{aligned}
        &
        \frac{\partial_i \vect{A}\cdot \partial_j \vect{A}}{1- \vect A^2}+\frac{\partial_i  \vect A^2 \partial_j \vect A^2 }{4(1- \vect A^2)^2}
        =
        \lambda_1 \delta_{ij}
        \ ,
        \\
        &
        \frac{\partial_i \vect{A}\cdot \partial_j \vect{B}+\partial_j \vect{A}\cdot \partial_i \vect{B}}{\sqrt{1-\vect A ^2}}
        = 
        \lambda_2 \delta_{ij}
        \ ,
        \\
        &
        \partial_i \vect{B}\cdot \partial_j \vect{B}-(\vect{A}\cdot \partial_i \vect{B})(\vect{A}\cdot \partial_j \vect{B})
        =
        \lambda_3 \delta_{ij}
        \ .
    \end{aligned}
\end{equation}
Each of the above equations is a nonlinear system of PDEs for the vectors $A^n$ and $B^n$. A way to simplify the situation is to assume the following dependence for $B^n$:
\begin{equation}\label{eq:ansatz_Bi}
    \vect{B}=k\vect{A} \ f(\vect A^2) \ ,
\end{equation}
where $k$ is a constant and $f$ a function to be determined.
Plugging this relation back inside the second and third equations of the previous system we obtain
\begin{equation}
\begin{aligned}
    &
    \frac{2k f \, \partial_i \vect{A} \cdot \partial_j \vect{A}}{\sqrt{1-\vect A^2}}+ \frac{kf' \partial_i \vect A^2 \partial_j \vect A^2}{\sqrt{1-\vect A^2}}
    =
    \lambda_2 \delta_{ij}
    \ , 
    \\
    &
    k^2 f^2 \partial_i \vect{A}\partial_j \vect{A}
    + k^2 \partial_i  \vect A^2 \partial_j \vect A^2
    \left[
    f f'+f'^2 \vect A^2-\left(\frac{f}{2}+f' \vect A^2\right)^2
    \right]
    =
    \lambda_3 \delta_{ij}
    \ .
    \end{aligned}
\end{equation}
We can notice that all these equations contain two contraction structures, $\partial_i \vect A \cdot \partial_j \vect A$ and $\partial_i \vect A^2 \partial_j \vect A^2$. Therefore, we can impose the coefficients of these structures to match the ones appearing in the first equation of \eqref{eq:sys_appA}. These conditions yield the following solutions
\begin{equation}
    f(\vect A^2)
    =
    \frac{1}{2\sqrt{1-\vect A^2}}
    \ , \quad 
    \lambda_2 
    =
    k\lambda_1
    \ , \quad
    \lambda_3
    =
    \frac{k^2}{4}\lambda_1
    \ .
\end{equation}
As a consequence, the system of nonlinear PDEs reduces to a single nonlinear PDE for the vector $\vect{A}$ alone,
\begin{equation}
    \frac{\partial_i \vect{A}\cdot \partial_j \vect{A}}{1-\vect A^2}
    +
    \frac{\partial_i \vect A^2 \partial_j \vect A^2 }{4(1-\vect A^2)^2}
    =
    \lambda_1 \delta_{ij}
    \ ,
\end{equation}
enabling to write $(X^2)^i_j$ as (see last line of Eq.~\eqref{setagain})
\begin{equation}
      a^2G^2\delta_{ij}
      = 
      \lambda_1 
      \left( \frac{k}{2}+\sigma_u \int_{t_0}^t \de t' \ F(t') \right)^2 
      \delta_{ij}
      \ ,
\end{equation}
which is the relation\footnote{To be precise, the relation used in the main text is obtained identifying $\sigma$ with $\frac{k}{2}\sigma_u$.} we used in Sec.~\ref{sec:dRGT_bkg} to fully solve the background cosmology and find the time dependence of $F$ and $G$.
Notice that, given the knowledge of $\vect{B}$ from Eq.~\eqref{eq:ansatz_Bi}, one can also determine $C$ as
\begin{equation}
    \partial_i C
    =
    \frac{k \partial_i \vect A^2}{4{(1-\vect A^2)}^{3/2}}
    \quad
    \Longrightarrow 
    \quad 
    C = k f(\vect A^2)+k_0 \ .
\end{equation}
As a consequence, the family of viable Stueckelberg configurations is now parametrized by a vector $A^n$ and some real parameters:
\begin{equation}
	\phi^0
	=
	\sigma_u \frac{\int_{t_0}^t \de t' \, F({t'})}{\sqrt{1-\vect{A}^2}}+ k f(\vect A^2)+k_0
	\quad 
	\Longrightarrow
	\quad
	\phi^n
	= 
	A^n\left[\sigma_u\frac{\int_{t_0}^t \de t' \, F({t'})}{\sqrt{1-\vect{A}^2}}+k f(\vect A^2)\right]
	\ .
\end{equation}

\subsection{Mixed Branch}

The Mixed Branch cosmology is characterized by $F=0$ and $G=g/a$ (with $g$ constant). As a consequence, the system in Eq.~\eqref{setagain} reduces to
\begin{equation}\label{eq:systemMix}
    \begin{cases}
	0
	=
	(\dot{\phi}^0)^2(1 - \vect A^2)
	\ ,
	\\
    \dot{\phi^0}
    (\partial_i \phi^0-A^n \partial_i \phi^n)
    =
    0
    \ ,
    \\
	g^2\delta_{ij}
	=
	-\partial_i \phi^0 \partial_j \phi^0
	+ \partial_i \phi^n \partial_j\phi^n 
	\ ,
    \end{cases}
\end{equation}
with $g$ a constant. As a consequence, the admitted profiles are either static configurations or configurations for which $A^n$ has unit norm. Let us study these two cases separately.

\subsubsection{$\dot{\phi}^0=0$}

When $\dot{\phi}^0=0$ the first two equations of \eqref{eq:systemMix} are automatically satisfied. Moreover, the condition that characterizes this branch (i.e.~$F=0$) implies that the Stueckelberg fields must be time-independent. Notice that, when $F=0$, the EOM\footnote{This can be verified by varying the dRGT Action wrt $\phi^a$ keeping in mind that the dRGT potentials now only depend on $G$.} of the Stueckelberg fields imply that each $\phi^a$ must be a harmonic function:
\begin{equation}
    \nabla^2\phi^a = 0 \ .
\end{equation}
It can be easily verified that a simple solution of the system above is given by
\begin{equation}
    \phi^a=C^a+B^a_ix^i \ ,
\end{equation}
for any constant matrix $B^a_i$ which satisfies 
\begin{equation}
    B^a_i B^b_j\eta_{ab}=g^2\delta_{ij} \ .
\end{equation}
A family of viable profiles is given by the equivalence class of profiles $\{\Lambda^a_b\bar{\phi}^b\}$; with $\bar{\phi}^0=0$, $\bar{\phi}^i\propto x^i$ being the configuration found in \cite{Volkov:2012zb}.

\subsubsection{$\vect{A}^2=1$}

In this last case, it is still possible to relate $\phi^n$ to $\phi^0$. Indeed, using the second equation of \eqref{eq:systemMix} we obtain
\begin{equation}
    \phi^n
    = 
    A^n \phi^0+B^n
    \>,\> 
    \vect{A}\cdot \partial_i \vect{B}=0
    \ .
\end{equation}
Moreover, from the third equation of Eq.~\eqref{eq:systemMix} $ (X^2)^i_j$ becomes
\begin{equation}
    g^2 \delta_{ij}
    =
    \partial_i \vect{B}\cdot \partial_j \vect{B}
    +\left( \partial_i \vect{A}\cdot \partial_j \vect{B}+\partial_j \vect{A}\cdot \partial_i \vect{B}\right)\phi^0
    +\partial_i \vect{A}\cdot \partial_j \vect{A}(\phi^0)^2
    \ .
\end{equation}
Similarly to what happened for the $\Lambda$-Branch, we must look for diagonal solutions of the kind
\begin{equation}
    \partial_i \vect{B}\cdot \partial_j \vect{B}
    =
    h_i(\vect x)\delta_{ij}
    \>
    ,
    \quad
    \partial_i \vect{A}\cdot \partial_j \vect{B}
    +\partial_j \vect{A}\cdot \partial_i \vect{B}
    =
    l_i(\vect x)\delta_{ij}
    \>
    ,
    \quad
    \partial_i \vect{A}\cdot \partial_j \vect A
    =
    m_i(\vect x) \delta_{ij}
    \ ,
\end{equation}
where, with abuse of notation, we are indicating that each of the above matrices must be diagonal but their entries $h_i(\vect x)$, $l_i(\vect x)$, $m_i(\vect x)$ can be different functions of the spatial coordinates. This freedom comes from the fact that each of these matrices is multiplied by a different power of $\phi^0(t,\vect x)$.

\section{Perturbations with Matter}\label{apppert}

To study the stability of metric perturbations in the presence of a generic fluid it is convenient to study the linearized Einstein equations. Similarly to Sec.~\ref{sec:perturbs}, we work in unitary gauge and split the metric tensor as $g_{\mu \nu}=\bar{g}_{\mu \nu}+h_{\mu \nu}$, where the components of the perturbations are given by Eq.~\eqref{eq:SVT}, while the stress-energy tensor of the matter sector is given by
\begin{equation}
\begin{aligned}
    &
    T_{00}
    =
    \bar{\rho}+\delta \rho-A\bar{\rho}
    \>,\quad 
    T_{i0}
    =
    -\bar{\rho}(\hat{B}_i+\partial_i B)-a^2(\bar{\rho}+\bar{p})(\hat{v}_i+\partial_i v) 
    \>,
    \\ 
    & T_{ij}
    =
    a^2\delta_{ij} (\bar{p}+\delta p)+\partial_{\langle i}\partial_{j\rangle}\Pi + \partial_{(i}\hat{\Pi}_{j)}+\hat{\Pi}_{ij}
    \ .
\end{aligned}    
\end{equation}
We will also need to consider the linearized conservation equations for both the Stueckelberg and Matter Stress-Energy tensors:
\begin{equation}
	\nabla_{\mu}Y^{\mu}_{\nu} = 0
	\>,
	\quad
	\nabla_{\mu}T^{\mu}_{\nu}=0 \ .
\end{equation}

\subsection{\texorpdfstring{$\Lambda$}{Lambda}-Branch}

On the $\Lambda$-branch, the conservation equation of $Y^{\mu}_{\nu}$ acquires the following form
\begin{equation}
    \nabla_{\mu}Y^{\mu}_{0}
    =
    \mathcal{F}_C C=0
    \>,
	\quad
    \nabla_{\mu}Y^{\mu}_{i}
    =
    \mathcal{F}_E \partial_i \nabla^2 E=0
    \>,
	\quad
    \nabla_{\mu}Y^{\mu}_{i}
    =
    \mathcal{F}_{\hat{E}_i} \hat{E}_i=0
    \ ,
\end{equation}
where $\mathcal{F}_{\mathcal{O}}$ are functions of time, for $\mathcal{O}= \{ C, E, \hat{E}_i \}$. As a consequence, the two scalars $C$, $E$ and the vector $\hat{E}_i$ are forced to vanish. Let us now explore the behaviour of each perturbative sector.
\subsubsection{Scalars}
The scalar content of the Einstein equations consists of 
\begin{equation}\label{eq:Einstein_matter_pert}
    \begin{aligned}
	&
	-2 \frac{H}{a^2}\nabla^2 B
	+\frac{3H^2}{\MP^2}A
	=
	\frac{\delta \rho}{\MP^2}
	\ ,
	\\
	&
	\partial_i 
	\left[
	-2\dot{H}B-H A +a^2\frac{\bar{\rho}_m+\bar{p}_m}{\MP^2}  v
	\right]
	=
	0
	\ ,
	\\
	&
	\partial_{\langle i}\partial_{j \rangle}
	\bigg[ 
	\frac{1}{2} A- H B- \dot{B}-a^2\frac{\Pi}{\MP^2}
	\bigg]
	=
	0
	\ ,
	\\
	&
	-\nabla^2 A -3a^2(3H^2+2\dot{H})A-3a^2H\dot{A}+2\nabla^2(HB+\dot{B})=3a^2\frac{\delta p}{\MP^2}
	\ .
	\\
    \end{aligned}
\end{equation}
It is evident that neither $A$ nor $B$ are dynamical metric perturbations. Indeed, the first two equations of \eqref{eq:Einstein_matter_pert} (respectively, the $00$ and $0i$ components of the Einstein equations) can be used to fully determine the remaining metric perturbations in terms of $\delta \rho$ and $v$. It can be concluded that the scalar sector is plagued by strong coupling, as no degree of freedom is able to propagate at the linear level.

\subsubsection{Vectors}

The vector part of the $0i$ and $ij$ components of the Einstein equations is given by
\begin{equation}
    \begin{aligned}
	&
	-\frac{m^2 f_B}{2}\hat{B}_i -\left(3H^2+2\dot{H}\right)\hat{B}_i-\frac{1}{2a^2}\nabla^2 \hat{B}_i
	=
	-a^2\frac{\bar{\rho}_m+\bar{p}_m}{\MP^2}\hat{v}_i - \frac{\bar{\rho}_m}{\MP^2} \hat{B}_i
	\ ,
	\\
	&
	- H\partial_{(i}\hat{B}_{j)}-\partial_{(i}\dot{\hat{B}}_{j)}
	=
	\frac{\partial_{(i}\hat{\Pi}_{j)}}{\MP^2}
	\ ,
    \end{aligned}
\end{equation}
where $f_B$ is a function of $F(t)$ and of the parameters $\alpha_i$. Since $\hat{B}_i$ can be algebraically determined in terms of the matter perturbations we can see that also the vector sector is strongly coupled, as no metric degree of freedom propagates on the $\Lambda$-Branch.

\subsubsection{Tensors}

The evolution of the tensor modes\footnote{Notice that $\hat{h}_{ij}$ differs from its main text counterpart by a factor $a^2$.} is governed by 
\begin{equation}
   \ddot{\hat{h}}_{ij}+3H\dot{\hat{h}}_{ij}- \frac{1}{a^2}\nabla^2 \hat{h}_{ij}+\left( 2\frac{\bar{p}_{\phi}}{\MP^2}-m^2 f_h\right)\hat{h}_{ij}
   =
   2\frac{\hat{\Pi}_{ij}}{\MP^2}
   \ ,
\end{equation}
where $f_h$ depends on $F(t)$ and on the dRGT parameters. Similarly to the discussion of Sec.~\ref{sec:perturbs}, the tensor modes are the only dynamical degrees of freedom on the $\Lambda$-Branch. Their evolution equation differs from its GR form by the addition of a mass term (remember that $\bar{p}_{\phi}\propto m^2 \MP^2$), as expected for a massive degree of freedom.

\subsection{Mixed Branch}

On the Mixed Branch, the conservation equation of $Y^{\mu}_{\nu}$ acquires the form
\begin{equation}
    \begin{aligned}
    &
    \nabla_{\mu}Y^{\mu}_0 = 0
    \ ,
    \\
    &
    \nabla_{\mu}Y^{\mu}_i
    =
    3\beta_0 A-6(2\tilde{y}HB+\beta_0 \dot{B})+2\tilde{y}(\nabla^2 \tilde{E}-3\tilde{C})
    =
    0
    \ ,
    \\
    &
    \nabla_{\mu}Y^{\mu}_i=\tilde{y}\frac{\nabla^2 \hat{E}_i}{a^2}-4(\beta_0 \dot{\hat{B}}_i+2\tilde{y}H\hat{B}_i)
    =
    0
    \ ,
    \end{aligned}
\end{equation}
where $\tilde{y}$ and $\beta_0$ are functions of the scale factor and the dRGT parameters and obey
\begin{equation}
    \dot{\beta}_0 + 2H(\beta_0-\tilde{y}) = 0 \ .
\end{equation}
Notice that the first equation is identically satisfied. 

\subsubsection{Scalars}

The full set of Einstein equations for the scalar sector
is given by
\begin{equation}
    \begin{aligned}
	&
	-2 \frac{H}{a^2}\nabla^2 B
	+3 H \dot{\tilde{C}}
	-\frac{\nabla^2\tilde{C}}{a^2}
	+\frac{1}{3a^2}\nabla^4 \tilde{E}
	+3\frac{m^2G \beta_0}{4}\tilde{C}
	+3H^2A
	=
	\frac{\delta \rho}{\MP^2}
	\ ,
	\\
	&
	\partial_i 
	\left[
	-2\dot{H}B-H A -\dot{\tilde{C}}+\frac{1}{3}\nabla^2 \dot{\tilde{E}}
	\right]
	=
	-a^2\frac{\bar{\rho}_m+\bar{p}_m}{\MP^2} \partial_i v
	\ ,
	\\
	&
	\partial_{\langle i}\partial_{j \rangle}
	\bigg[
	\frac{1}{2} A- H B- \dot{B} 
	- \frac{1}{2} \tilde{C}
	+\frac{a^2}{2}
	\left(\ddot{\tilde{E}}+3H\dot{\tilde{E}}+\left(\frac{\bar{p}_{\phi}}{\MP^2}-m^2\tilde{f}_h\right)\tilde{E}\right)
	+\frac{1}{6} \nabla^2 \tilde{E}
	\bigg]
	=
	a^2\frac{\partial_{\langle ij \rangle}\Pi}{\MP^2}
	\ ,
	\\
	&
	-\nabla^2 A 
	-3a^2(3H^2+2\dot{H})A
	-3a^2H\dot{A}+2\nabla^2(HB+\dot{B})
	-3a^2
	\left[
	\ddot{\tilde{C}}+3H\dot{\tilde{C}}
	-\left(\frac{\bar{p}_{\phi}}{\MP^2}
	-\frac{m^2 \tilde{f}_C}{2}\right)\tilde{C}
	\right]
	+
	\\
	&
	+\nabla^2\tilde{C}-\frac{1}{3}\nabla^4 \tilde{E}
	=
	\frac{3a^2 \delta p}{\MP^2}
	\ ,
	\\
 \end{aligned}
\end{equation}
where $\tilde{f}_h$ and $\tilde{f}_C$ depend on the scale factor as well as on the dRGT parameters.
On the other hand, the conservation of the matter stress-energy tensor yields
\begin{equation}
    \begin{aligned}
	&
	\nabla_{\mu} T^{\mu}_0 
	\ : \quad
	-\frac{3}{2}(\bar{\rho}_m+\bar{p}_m)\dot{\tilde{C}}
	-2a^2
	\left[
	3H(\delta p+\delta \rho)+\delta\dot{\rho}+(\bar{\rho}_m+\bar{p}_m)\nabla^2 v
	\right]
	=
	0
	\ ,
	\\
	&
	\nabla_{\mu} T^{\mu}_i 
	\ : \quad
	-\frac{1}{2}(\bar{p}_m+\bar{\rho}_m)A+\frac{1}{a^3}
	\partial_t
	\left[a^3(\bar{p}_m+\bar{\rho}_m)(B+a^2 v)\right]
	+\delta p
	+\frac{2}{3}\nabla^2 \Pi
	=
	0
	\ .
    \end{aligned}
\end{equation}
First of all, the continuity equation can be integrated to fully express $\tilde{C}$ in terms of the matter perturbations. Moreover, the $00$ and $0i$ components of the Einstein equations can be solved to yield $A[\tilde{E},\dot{\tilde{E}};\delta_{\mathrm{matter}}]$ and $B[\tilde{E},\dot{\tilde{E}};\delta_{\mathrm{matter}}]$. Thus, we can combine the trace-free and trace parts of Einstein equations to extract an evolution equation for the scalar $\tilde{E}$. The detailed expression is complicated and not very illuminating, but we can schematically write it, in Fourier space, as
\begin{equation}\label{eq:pertES}
    \ddot{\tilde{E}}_k
    +\mathcal{G}_1(k,t)\dot{\tilde{E}}_k
    +\mathcal{G}_2(k,t)\tilde{E}_k
    =
    \mathcal{S}_{\rm S}[\delta_{\mathrm{matter,} k}]
    \ .
\end{equation}
The source term on the RHS is built out of matter perturbations, while the LHS is the differential operator which captures the time evolution of the scalar modes $\tilde{E}_k$. To verify whether the addition of matter succeeds in curing the pathologies encountered in Sec.~\ref{sec:perturbs}, we take the $k\to \infty$ limit of the above equation. In this limit, the LHS simplifies to
\begin{equation}
    \ddot{\tilde{E}}_k+\mathcal{G}_0 \tilde{E}_k+\mathcal{O}\left(\frac{H}{k}\right) \ .
\end{equation}
Thus, in the sub-horizon limit, we end up with a zero speed of sound for the scalar mode. We can therefore conclude that the presence of ordinary matter does not affect the qualitative behaviour of the scalar sector, which is still plagued by strong coupling. However, the presence of scalar anisotropic stress would bring contributions proportional to $\tilde{E}_k$ in the RHS of eq.~\eqref{eq:pertES}.

\subsubsection{Vectors}

The vector content of the Einstein equations is given by
\begin{equation}
    \begin{aligned}
	& 
	-2\dot{H}\hat{B}_i-\frac{1}{2a^2}\nabla^2 \hat{B}_i-\frac{H}{2}\nabla^2 \hat{E}_i
	=
	-a^2\frac{\bar{\rho}_m+\bar{p}_m}{\MP^2}\hat{v}_i
	\ ,
	\\
	&
	- H\partial_{(i}\hat{B}_{j)}
	-\partial_{(i}\dot{\hat{B}}_{j)}
	+\frac{a^2}{2}\partial_{(i}
	\left[
	\left(\frac{\bar{\rho}_{\phi}}{\MP^2}-m^2\tilde{f}_h\right) 
	\hat{E}_{j)}+3H\dot{\hat{E}}_{j)}+ \ddot{\hat{E}}_{j)}
	\right]
	=
	a^2\frac{\partial_{(i}\hat{\Pi}_{j)}}{\MP^2}
	\ .
    \end{aligned}
\end{equation}
Differently from the $\Lambda$-Branch, the vector $\hat{E}_i$ is now able (in principle) to propagate. Combining the two equations (expressing, first of all, $\hat{B}_i$ in terms of $\hat{E}_i$ and the matter perturbations) enables us to find an evolution equation for the vector d.o.f. $\hat{E}_i$:
\begin{equation}
    \ddot{\hat{E}}_{i,k}+\mathcal{V}_1(k,t)\dot{\hat{E}}_{i,k}+\mathcal{V}_2(k,t)\hat{E}_{i,k}
    =
    \mathcal{S}_{\rm V}[\delta_{\mathrm{matter},k}]
    \ .
\end{equation}
Along the lines of the previous subsection, we are not interested in the details of the form of the matter source or the differential operator on the LHS. Rather, we will simply extract the shape of the LHS in the subhorizon limit, which happens to be
\begin{equation}
    \ddot{\hat{E}}_i+\frac{k^2\tilde{y}}{2a^2\beta_0}\hat{E}_i+\mathcal{O}\left(\frac{H}{k}\right).
\end{equation}
Thus, contrary to the scalar sector, the vector mode keeps propagating on sub-horizon scales, with a speed of sound 
\begin{equation}
    c_s^2=\frac{\tilde{y}}{2\beta_0}.
\end{equation}
A similar conclusion was obtained in Sec.~\ref{sec:perturbs}.

\subsubsection{Tensors}

On the Mixed Branch, the tensor modes evolve according to
\begin{equation}
     \ddot{\hat{h}}_{ij}
     +3H\dot{\hat{h}}_{ij}
     - \frac{1}{a^2}\nabla^2 \hat{h}_{ij}
     +\left( 2\frac{\bar{p}_{\phi}}{\MP^2}
     -m^2 \tilde{f}_h\right)\hat{h}_{ij}
     =
     2\frac{\hat{\Pi}_{ij}}{\MP^2}
     \ .
\end{equation}
Once again, the tensor modes are free to propagate on the background, with an evolution equation which deviates from its standard GR form by the addition of the matter term. In the large-$k$ limit, the tensor modes propagate with a speed of sound $c_s^2=1$.

\bibliographystyle{utphys}
\bibliography{biblio}

\end{document}